\documentclass[12pt]{article}
\usepackage{graphicx,amssymb,amsfonts,amsmath,amsthm,xspace,color,ulem,url}
\usepackage{scicite,times}

\newenvironment{sciabstract}{%
\begin{quote} \bf}
{\end{quote}}

\newcounter{lastnote}

\title{Clique topology reveals intrinsic geometric structure in neural correlations} 

\author
{Chad Giusti$^{2,3}$, Eva Pastalkova$^{4}$, Carina Curto,$^{1,3\dagger}$ Vladimir Itskov$^{1,3\dagger\ast}$\\
\\
\normalsize{$^{1}$Department of Mathematics \& Center for Neural Engineering, 
The Pennsylvania State University}\\
\normalsize{$^{2}$Warren Center for Network and Data Sciences, 
University of Pennsylvania}\\
\normalsize{$^{3}$Department of Mathematics, 
University of Nebraska--Lincoln}\\
\normalsize{$^{4}$Janelia  Research Campus, Howard Hughes Medical Institute, Ashburn, VA 20147, USA}\\
\normalsize{$\dagger$ These authors contributed equally to this work.}
\\
\normalsize{$^\ast$To whom correspondence should be addressed:  vladimir.itskov@psu.edu.}
}

\date{}

\begin{document} 
\baselineskip24pt

\baselineskip15pt 
\def\ord{\operatorname{ord}}


 \maketitle

\begin{sciabstract}
Detecting meaningful structure in neural activity and connectivity data is challenging in the presence of hidden nonlinearities, where traditional eigenvalue-based methods may be misleading.  We introduce a novel approach to matrix analysis, called clique topology, that extracts features of the data invariant under nonlinear monotone transformations.  These features can be used to detect both random and geometric structure, and depend only on the relative ordering of matrix entries. 
We then analyzed the activity of pyramidal neurons in rat hippocampus, recorded while the animal was exploring a two-dimensional environment, and confirmed that our method is able to detect geometric organization using only the intrinsic pattern of neural correlations.  Remarkably, we found similar results during non-spatial behaviors such as wheel running and REM sleep.  This suggests that the geometric structure of correlations is shaped by the underlying hippocampal circuits, and is not merely a consequence of position coding.  We propose that clique topology is a powerful new tool for matrix analysis in biological settings, where the relationship of observed quantities to more meaningful variables is often nonlinear and unknown.
\end{sciabstract}


\pagebreak

Neural activity and connectivity data are often presented in  the form of a matrix whose entries, $C_{ij},$ indicate the strength of correlation or connectivity between pairs of neurons, cell types, or imaging voxels.  
Detecting structure in such a matrix is a critical step towards understanding the organization and function of the underlying neural circuits.  This structure may reflect the coding properties of neurons, rather than their physical locations within the brain.  
For example, pyramidal neurons in rodent hippocampus possess a geometric organization due to their role in position coding.  Each of these neurons, called {\it place cells}, acts as a position sensor, exhibiting a high firing rate when the animal's position lies inside the neuron's {\it place field}, its preferred region of the spatial environment \cite{OKeefe1971}.  Because of this organization, the pairwise correlations $C_{ij}$ between place cells decrease as a function of the distances between place field centers \cite{Hampson1996}, and the matrix of correlations thus inherits a geometric structure.  
Alternatively, a correlation or connectivity matrix could be truly unstructured, such as the connectivity pattern in the fly olfactory system observed in \cite{Caron2013}, indicating random network organization.  

Can we detect the presence of structure -- or randomness -- purely from the intrinsic features of a matrix $C_{ij}$?  The most common approach is to use standard tools from matrix analysis that rely on quantities, such as eigenvalues, that are invariant under linear change of basis.  This strategy is natural in physics, where meaningful quantities should be preserved
by linear coordinate transformations.  
In contrast, structure in neural data should be invariant under matrix transformations of the form
\begin{equation}\label{eqn:nonlinear-transform}
C_{ij}=f(A_{ij}),
\end{equation}
where $f$ is a monotonically increasing function (Figure 1a).  This is because it is common for observed quantities in neuroscience to be related to more meaningful variables by a monotonic nonlinearity, such as the relationship between the strength of an imaging signal and the underlying neural activity \cite{calcium-imaging2012}, or the relationship between neural activity and represented stimuli.  For example, 
in the case of hippocampal place cells, pairwise correlations decrease with distance between place field centers in a nonlinear fashion.  In less-studied contexts, the relationship of neural correlations to the represented stimuli may be completely unknown.

Unfortunately, eigenvalues are {\it not} invariant under transformations of the form~\eqref{eqn:nonlinear-transform} (e.g. Supplementary Figure 1).  Although large random matrices have a reliable eigenvalue spectrum (e.g. Wigner's semicircle law \cite{Wigner1958}),
it is possible that a random matrix with i.i.d. entries could be mistaken as structured, purely as an artifact of a monotonic nonlinearity (Figure 1b).\footnote{Note that the matrix $C=f(A)$ in Figure 1b is also a random matrix, with i.i.d. entries drawn from the transformed distribution.  Although the eigenvalue spectrum still converges to the Wigner semicircle distribution in the limit of large $N$ \cite{PasturShcherbina2011}, the rate of convergence depends on the nonlinearity $f$, allowing for large deviations from the semicircle distribution as compared to a normally-distributed random matrix with the same $N$.}  The results of eigenvalue-based analyses can thus be difficult to interpret, and potentially misleading.

\begin{figure}[!h]\label{fig1}
\begin{center}
\includegraphics[width=6.3in]{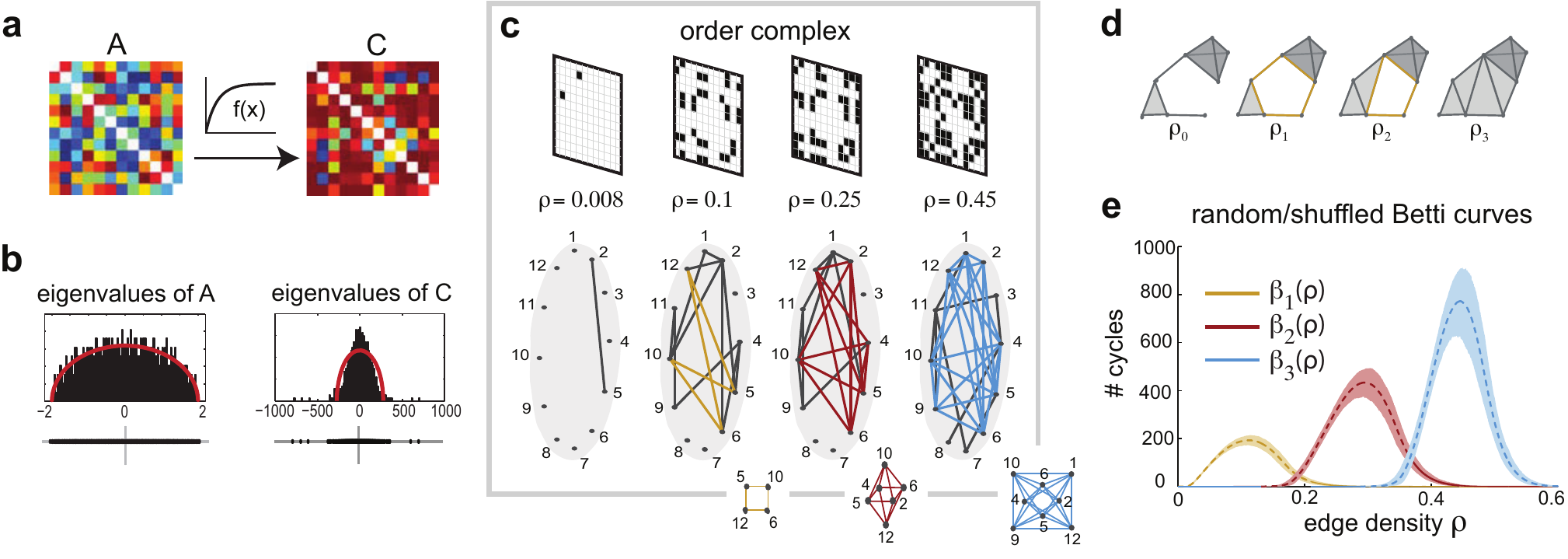}
\end{center}

\caption{\small {\bf Order-based analysis of symmetric matrices.}
(\textbf{a}) 
 A symmetric matrix $A$ is related to another matrix $C$ via a nonlinear monotonically  increasing function $f(x)$, applied entrywise. 
(\textbf{b})   (Left)
Distribution of eigenvalues  for a random symmetric $N \times N$ matrix $A,$  whose entries were drawn independently from the  normal distribution with zero mean and variance $1/\sqrt{N}$ ($N = 500$).
(Right) Distribution of eigenvalues for the transformed matrix with entries $C_{ij} = f(A_{ij})$, for $f(x)=1-e^{-30x}$.  Red curves show Wigner's semicircle distribution with matching mean and variance.
(\textbf{c})  (Top) The order complex of $A$ is represented as a sequence of binary adjacency matrices, indexed by the density $\rho$ of non-zero entries.   (Bottom) Graphs corresponding to the adjacency matrices.   Minimal examples of a $1$-cycle (yellow square), a $2$-cycle (red octahedron) and a $3$-cycle (blue orthoplex) appear at $\rho = 0.1, 0.25,$ and $0.45$, respectively.  
(\textbf{d}) At edge density $\rho_0$, there are no cycles.  Cliques of size 3 and 4 are depicted with light and dark gray shading.  As the edge density increases, a new $1$-cycle (yellow) is created, persists,  and is eventually destroyed at densities $\rho_1,\rho_2,$  and $\rho_3$, respectively.  
 (\textbf{e}) 
 For a distribution of 1000 random $N \times N$ symmetric matrices ($N = 88$), average Betti curves $\beta_ 1(\rho), \beta_2(\rho),$ and $\beta_3(\rho)$ are shown (yellow, red, and blue dashed curves), together with 95\% confidence intervals (shaded areas).   
}
\end{figure}

Here we introduce a new  tool to reliably detect signatures of structure and randomness that are invariant under transformations of the form~\eqref{eqn:nonlinear-transform}.  
The only feature of a matrix that is preserved under such transformations is the relative ordering of its entries, as $C_{ij} < C_{k\ell}$ whenever $A_{ij} < A_{k\ell}$ (see Supplementary Text).
We refer to this combinatorial information as the {\it order complex}, $\ord(C)$.  
It is convenient to represent the order complex as a nested sequence of graphs, where each subsequent graph includes an additional edge  $(ij)$ corresponding to the next-largest matrix entry $C_{ij}$ (Figure 1c).   Any quantity computed from the order complex is automatically invariant under the transformations~\eqref{eqn:nonlinear-transform}, since $\ord(A) = \ord(C)$.   

Perhaps surprisingly, the arrangement of cliques (all-to-all connected subgraphs) in the order complex of a matrix can be used in lieu of eigenvalues to detect random or geometric structure.  
{\it Clique topology} measures how cliques fit together and overlap. This can be quantified by counting non-contractible cycles, or ``holes,'' that remain after all cliques in a graph have been ``filled in'' (Figure 1c).    
As the edge density $\rho$ increases, new cycles are created, modified, and eventually destroyed (Figure 1d).  
  One can track these changes by computing a set of Betti numbers \cite{Hatcher,EdelsbrunnerHarer}, $\beta_m$, which count the independent  $m$-cycles.  The Betti numbers across all graphs in an order complex yield {\it Betti curves}, $\beta_m(\rho)$ (see Methods and Supplementary Text).   

Although the details of individual graphs in the order complex may be sensitive to noise in the matrix entries, we found that clique topology provides robust signatures of randomness.  
In the case of a random symmetric matrix with i.i.d. entries, the corresponding order complex is a sequence of Erdos-Renyi random graphs.  We found that the Betti curves $\beta_m(\rho)$ are remarkably reliable for such matrices (Figure 1e), and display a characteristic unimodal shape with peak values that increase with $m$ ($m \ll N$).  This reliability has been theoretically predicted \cite{Kahle2009, KahleMeckes2013}, and makes it possible to robustly distinguish random from non-random structure in the presence of a monotone nonlinearity~\eqref{eqn:nonlinear-transform}.  Unsurprisingly, correlation matrices obtained from finite samples of $N$ independent random variables display the same characteristic Betti curves as random symmetric $N\times N$ matrices (Supplementary Figure 2).  
Note that computing low-dimensional $(m \leq 3)$ Betti curves for matrices of size N$\sim$100 is numerically tractable due to recent advances in computational topology \cite{ZomorodianCarlsson2005,EdelsbrunnerHarer,HarkerMischaikowMrozekNanda2014}.  
\bigskip

\begin{figure}[!h]\label{fig2}
\begin{center}
\includegraphics[width=3in]{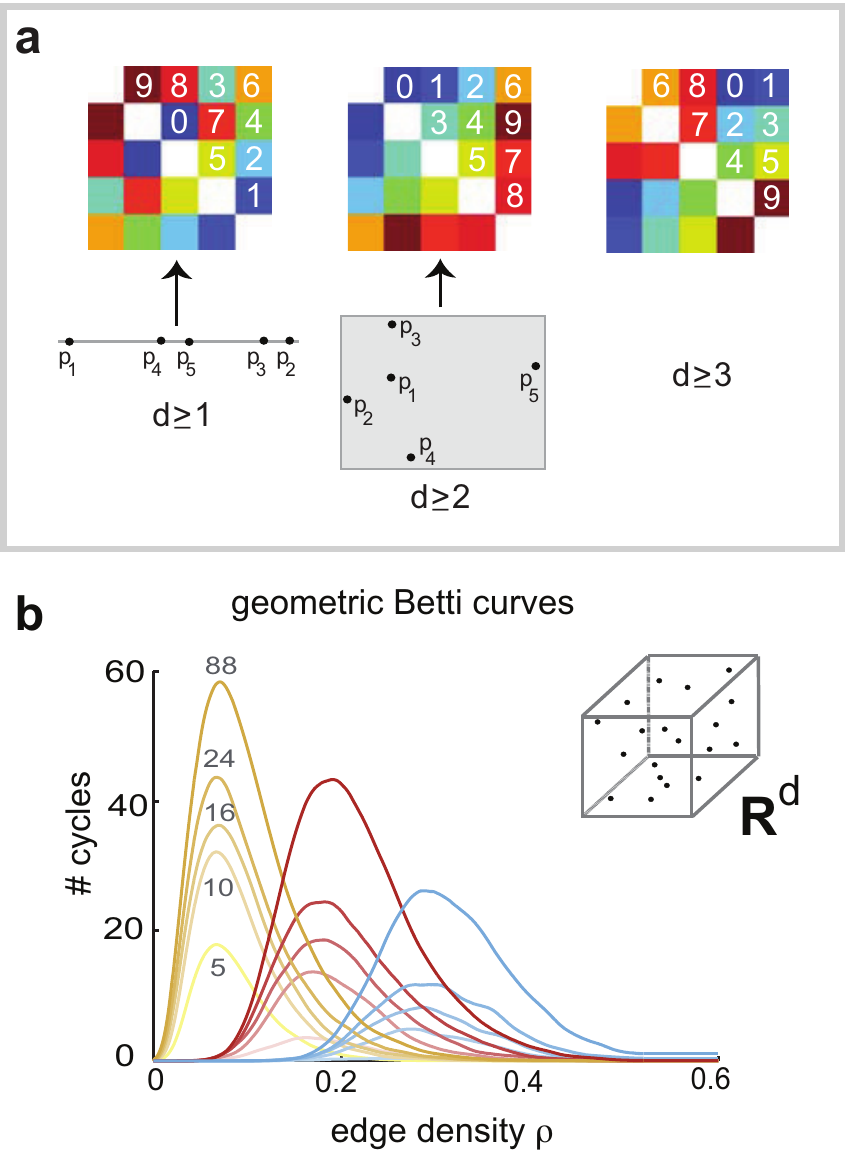}
\end{center}
\caption{\small {\bf Geometric structure is encoded in the ordering of matrix entries.}
(\textbf{a})  Three $5 \times 5$ symmetric matrices with distinct order complexes; the ten off-diagonal matrix values in each are ordered from 0 to 9.  (Left) An ordering of matrix values that can be obtained from an arrangement of points $p_i$ on a line, so that  $A_{ij} < A_{k\ell}$ whenever $||p_i - p_j|| < ||p_k - p_\ell||$.  (Middle) An ordering that arises from distances between points in the plane, but cannot be obtained from a one-dimensional point arrangement.   (Right) A matrix ordering that cannot arise from distances between points in one or two dimensions. 
 (\textbf{b}) Betti curves for distributions of geometric matrices ($N = 88$)
 in  dimensions $d = 5, 10, 16, 24,$ and $88.$  Mean 
Betti curves $\beta_ 1(\rho), \beta_2(\rho),$ and $\beta_3(\rho)$ are shown (yellow, red, and blue curves), with darker (and higher) curves corresponding to larger $d$. 
}
\end{figure}

If a correlation or connectivity matrix is not random, is there specific structure one can detect?  Uncovering {\it geometric} organization is especially important in the context of neuroscience.  Neurons tuned to features that lie in a continuous coding space, such as orientation-tuned neurons \cite{HUBELWIESEL1959} or hippocampal place cells \cite{OKeefe1971}, have correlations that decrease with distance.  
The geometry of the coding space may therefore be detectable in the pattern of pairwise correlations, even if the nonlinear relationship between distances and correlations is unknown.  Fortunately, the ordering of matrix entries encodes geometric features, such as dimension (Figure 2a).

For larger matrices, the precise dimension may be difficult to discern  in the presence of noise.  Nevertheless, the organization of cliques in the order complex carries signatures of an underlying Euclidean geometry, irrespective of dimension.  For example, the triangle inequality, $\Vert x-z \Vert \leq \Vert x-y\Vert + \Vert y-z \Vert$, implies that if two edges of a triangle are present at some edge density $\rho$, there  is a higher probability of the third edge also being present.  Intuitively, this means that cliques in the order complex will be more prominent for geometric as compared to random matrices, and cycles (Figure 1c,d) will be comparatively short-lived, as cliques cause holes to be more readily filled  \cite{Kahle2011}.

To see whether clique topology can provide reliable signatures of geometric organization, we computed Betti curves for distributions of geometric matrices ($N = 88$), generated from random points uniformly sampled from unit cubes of  dimensions $d = 5, 10, 16, 24,$ and $88,$ and having entries that {\it decrease} with distance (see Methods).
We then computed average Betti curves $\beta_1(\rho), \beta_2(\rho),$ and $\beta_3(\rho)$ for each $d$, and found that they are stratified by dimension but retain characteristic features that are independent of dimension ($d \leq N$).  In particular,
the peak values of geometric Betti curves are considerably smaller than those of random symmetric matrices with matching parameters ($p<0.001$), and {\it decrease} with increasing $m$ (Figure 2b).  
We conclude that  Betti curves can, in principle, be used to distinguish geometric from random structure.
\bigskip

Can clique topology be used to detect geometric organization from pairwise correlations in noisy neural data?   To answer this question, we examined correlations of hippocampal place cells in rodents during spatial navigation in a two-dimensional open field environment.  In this context, geometric structure is expected due to the existence of spatially localized receptive fields (place fields \cite{OKeefe1971}), but has not previously been detected {\it intrinsically} using only the pattern of correlations.
We computed correlations from spike trains of simultaneously recorded neurons in area CA1 of dorsal hippocampus (see Supplementary Methods).  Each pairwise correlation, $C_{ij},$ was obtained from the mean of a cross-correlogram on a timescale of $\tau_{\max} =1$s (see Methods; see also Supplementary Figure 3).  The resulting matrix was then analyzed using clique topology (Figure 3a).  As expected, the Betti curves from place cell data were in close agreement to those of geometric matrices (Figure 3b, top), up to a small rightward shift that is likely due to noise (Supplementary Figure 4).  
\bigskip

\begin{figure}[!h]\label{fig3}
\begin{center}
\includegraphics[width=6.3in]{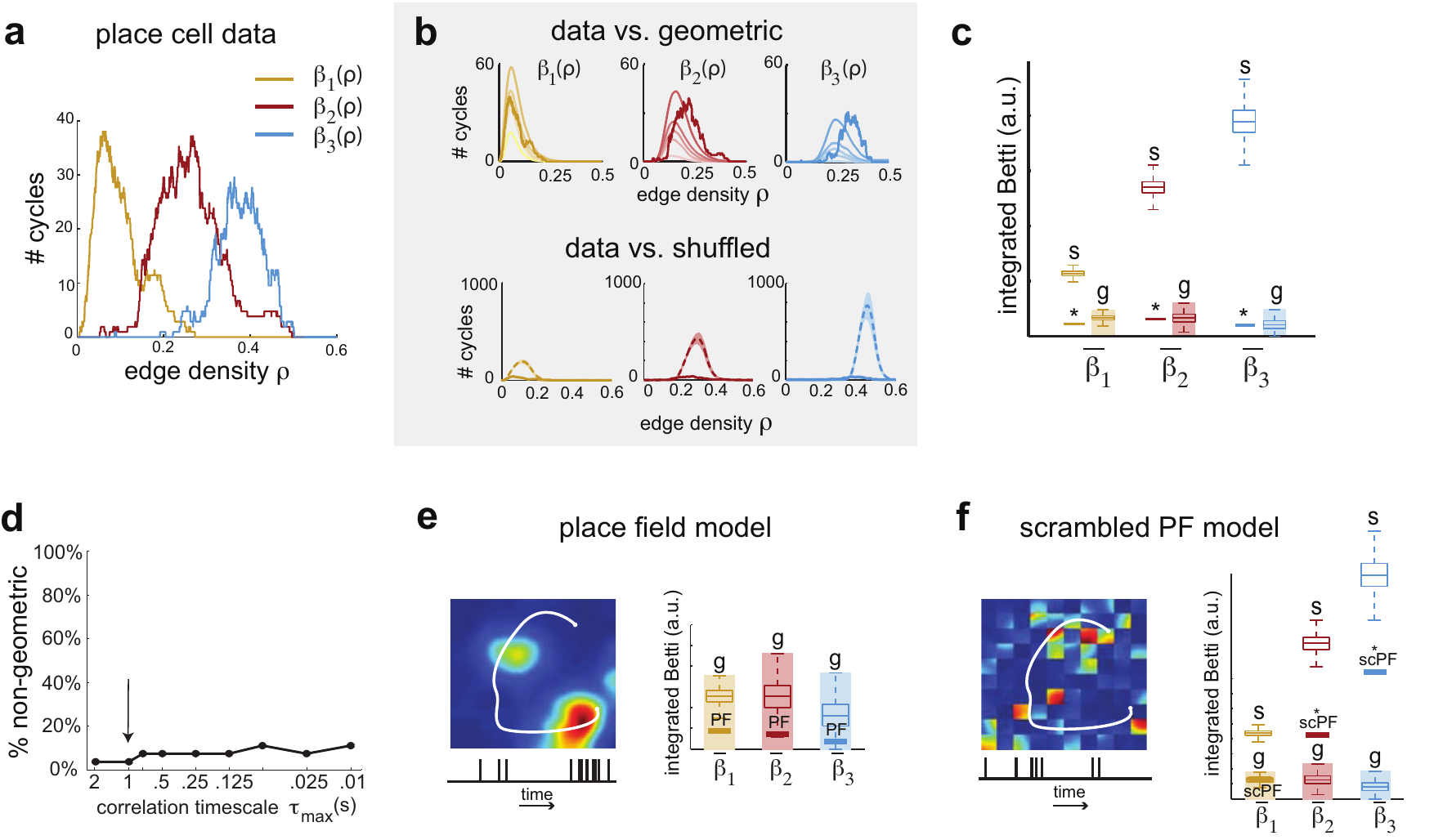}
\end{center}
\caption{\small \textbf{Geometric structure of correlations for neurons with spatial receptive fields.} 
(\textbf{a}) Betti curves of the pairwise correlation matrix for the activity of $N = 88$ place cells in hippocampus during  open field spatial exploration.   
(\textbf{b}) 
(Top) Betti curves from panel (a) (bold lines) overlaid on the mean geometric Betti curves from Figure 2b.
(Bottom) Comparison of Betti curves from panel (a)  to those of shuffled correlation matrices (note the change in vertical scale). 
(\textbf{c})  
Integrated Betti values $\bar \beta_m$ for the curves  in panel (a)  (solid lines), compared to standard box plots of integrated Betti values for the $1000$ shuffled [s] and geometric [g] controls displayed in (b).  The geometric box plots are shown for the highest dimension, $d = 88$, while the shaded area indicates the confidence interval across all dimensions $d\leq 88$.  Betti values for the place cell data are significantly non-random ($*$, $p<0.001$), and appear consistent with those of geometric matrices.
(\textbf{d}) Percentage of non-geometric Betti values  $\bar \beta_1$, $\bar \beta_2$, $\bar \beta_3$  for a range of correlation timescales $\tau_{\max}$.  Each point is an average over 9 open field recordings. The arrow indicates the timescale considered in (a)-(c). 
(\textbf{e}) (Left) A place field together with a cartoon trajectory (white) and simulated spike train (bottom).  
(Right) Integrated Betti values for correlations in the place field model (bold lines, labelled PF) lie within the geometric regime.  
(\textbf{f}) (Left) A scrambled version of the place field in (e).  
(Right)  Integrated Betti values for the scrambled PF model (bold lines) are significantly non-geometric ($*$, $p<0.05$) for $\bar{\beta}_2$ and $\bar{\beta}_3$, while $\bar{\beta}_1$ is in the geometric regime.  The Betti values are also significantly smaller than those of shuffled controls ($*$, $p<0.05$).  Box plots for geometric and shuffled controls in (e-f) are the same as in (c).
}
\end{figure}

We also compared the data Betti curves to shuffled controls, obtained by  randomly permuting the matrix entries (Supplementary  Figure 5a,b).  Shuffling completely destroys any structure in the order complex, yielding distributions of Betti curves identical to those of random matrices (Figure 1e).  We found that the Betti curves from place cell data were an order of magnitude smaller than the mean Betti curves of the shuffled matrices, and well outside the 95\% confidence intervals (Figure 3b, bottom). To quantify the significance of non-random structure, we used integrated the Betti values $$\bar \beta_m=\int_0^1 \beta_m(\rho) d\rho,$$  and verified that they were significantly smaller than those obtained from 1000 trials of the shuffled controls (p$<$0.001), but well within the confidence intervals for geometric controls (Figure 3c).  

To test whether the observed geometric organization was consistent across animals and recording sessions, we repeated these analyses for 8 additional data sets from three different animals during spatial navigation (Supplementary Figure 6).  All but one of the 9 data sets were consistent with the corresponding geometric controls, suggesting that geometric structure of correlations is a robust phenomenon during spatial navigation.  We also repeated the analyses for different choices of the correlation timescale, $\tau_{\max}$, ranging from $10$ ms to $2$ s, and observed similar results (Figure 3d).
 As a further test of geometric organization, we computed the distribution of {\it persistence lifetimes} from the order complex of the open field correlation matrix (see Supplementary Text).  The lifetime measures how long a hole persists as it evolves from one graph to the next in the order complex (Figure 1d).  Again, the data  exhibited topological signatures that were far from random, but consistent with geometric organization (Supplementary Figure 7).

To ensure that the observed correlation structure could not be explained by the differences in interactions of individual neurons with the ``mean field'' activity of the network, we performed an additional random control that preserves row and column sums of pairwise correlation matrices.  Specifically, we computed Betti curves for matrices drawn from a weighted maximum entropy (WME) distribution, subject to the constraint that expected row sums match the original pairwise correlation matrix (Supplementary Figure 5c,d).  The Betti curves and persistence lifetimes of the WME controls were similar to those of random symmetric matrices (Supplementary Figure 8), showing that the non-random structure in the data does not arise from the fact that some neurons have higher levels of correlation with the population as a whole.
 \bigskip

Are the spatial coding properties of place cells {\it sufficient} to account for the observed geometric organization of correlations during spatial navigation?  Or, alternatively, does this structure reflect finer features of the correlations, beyond what is expected from place fields alone?
To address this question, we computed place fields $F_i({\bf x})$ for each place cell from the same data used in Figure 3a, together with the animal's two-dimensional spatial trajectory ${\bf x}(t)$ (see Supplementary Methods).  We then generated synthetic spike trains for each neuron as inhomogeneous Poisson processes, with rate functions $r_i(t)$ given by the simple place field model
\begin{equation}\label{eqn:PFmodel}
r_i(t) = F_i\left({\bf x}\left (t\right)\right),
\end{equation} 
where ${\bf x}(t)$ is the animal's actual trajectory (see Figure 3e).
By design, the synthetic spike trains preserved the influence of place fields, but discarded all other features of the data, including precise spike timing and any non-spatial correlates.  Perhaps unsurprisingly, Betti curves derived from the place field model reproduced all the signatures of geometric organization (Figure 3e; see also Supplementary Figure 9b,c), indicating that place fields alone could account for the results observed in the open field data. 
 
We next asked whether the {\it geometry} of place fields was necessary, or if the Betti curves during spatial navigation could be attributed to an even more basic feature of the data, which is that each neuron is driven by the same global signal, ${\bf x}(t),$ filtered by a cell-specific function $F_i({\bf x})$.  To answer this question, we scrambled each place field by permuting the values of $F_i({\bf x})$ inside ``pixels''  of  a  $100\times 100$ grid, creating non-geometric receptive fields $\widetilde{F}_i({\bf x})$ (Supplementary Figure 9d; a $10 \times 10$ scrambling is shown in Figure 3f for clarity). We then generated spike trains from the actual trajectory, as in equation~\eqref{eqn:PFmodel}, but using the scrambled place fields $\widetilde{F}_i({\bf x})$.  For this model, we found that the second and third Betti curves were far outside of the geometric regime, while the first Betti curve $\beta_1(\rho)$ was insufficient to rule out geometric organization (Figure 3f; see also Supplementary Figure 9e-h). We obtained similar results after scrambling on a $10\times 10$ grid (Supplementary Figure 10).  We conclude that the geometric signatures observed during spatial navigation reflect the geometry of place fields, and are not simply a consequence of neurons being driven by the same global signal, ${\bf x}(t).$  Nevertheless, each of the Betti curves for the scrambled place field model was also significantly smaller than those of random controls (Figure 3f; p$<$0.001), suggesting that neurons controlled by a global signal via non-geometric receptive fields do exhibit non-random structure in their pairwise correlations.
\bigskip

\begin{figure}[!h]\label{fig4}
\begin{center}
\includegraphics[width=4in]{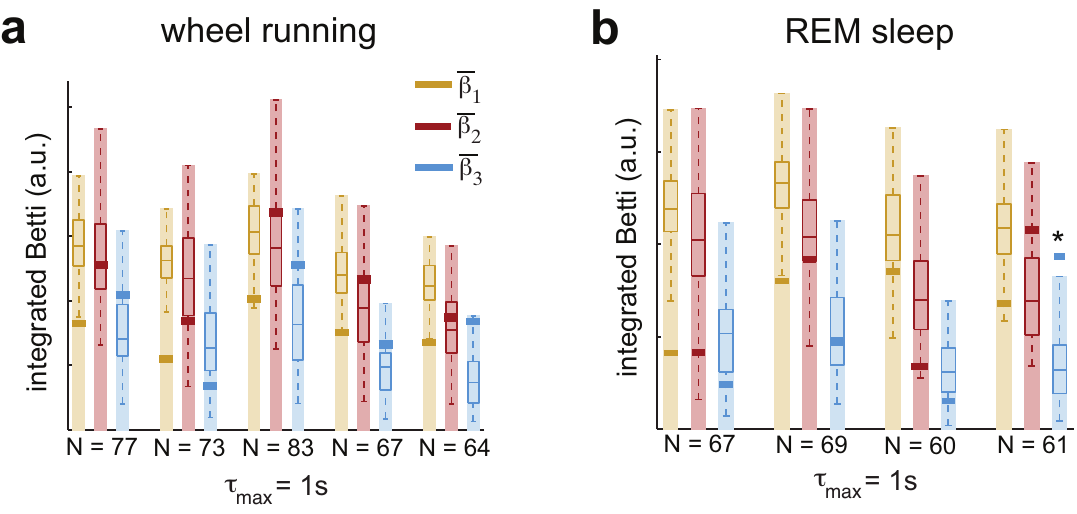}
\end{center}
\caption{\small {\bf Geometric organization in hippocampus during non-spatial behaviors.}
(\textbf{a}) Integrated Betti values $\bar \beta_1$, $\bar \beta_2$, and $\bar \beta_3$ (bold yellow, red, and blue lines) for 5 recordings  from 2 animals, during wheel running.  $N$ indicates the number of neurons in each recording.
Box plots indicate the distributions of Betti values for 100 geometric controls with  matching $N$ and dimension $d = N$.  Shaded regions indicate confidence intervals for the full geometric regime, with $d\leq N$. 
(\textbf{b})  Integrated Betti values for 4 recordings from 2 animals, during REM sleep.
 One Betti value was significantly non-geometric ($*$, $p < 0.05$).  
}
\end{figure}

The above results suggest that geometric structure in place cell correlations is a consequence of position coding, and is not necessarily expected during non-spatial behaviors.  To see if this is true, we repeated our analyses on neural activity recorded during two non-spatial conditions:  wheel running and REM sleep.  Surprisingly, we found that the Betti curves were again highly non-random (Supplementary Figure 11), and consistent with geometric organization across all 5 wheel running recordings and 3 out of 4 sleep recordings (Figure 4).  These findings suggest that geometric organization on a timescale of $\tau_{\mathrm{max}} \sim 1$s is a property of the underlying hippocampal network, and not merely a byproduct of spatially structured inputs.  At much finer timescales, however, geometric features appear to deteriorate in both REM sleep and wheel-running conditions (Supplementary Figure 12), in contrast to the open field data (Supplementary Figure 13).  

\paragraph{Discussion.}

We have developed a novel tool for detecting structural features of symmetric matrices that are invariant under nonlinear monotone transformations.  We have shown that this method can reliably detect both geometric and random structure in the presence of an unknown nonlinearity.  Our approach exploits the little-known fact that the ordering of  matrix entries, irrespective of their actual values, carries significant information about the underlying matrix organization.  Unlike eigenvalues, which can be badly distorted by monotone nonlinearities, the information encoded in the order complex is invariant.  Applying techniques from computational topology, relevant features can be extracted from the order complex that enable robust detection of geometric (or random) structure.  Unlike previous instances of topological data analysis \cite{gap, SinghMemoliIshkhanovSapiroCarlssonRingach2008, NicolauLevineCarlsson2011, DabaghianMemoliFrankCarlsson2012, ChanCarlssonRabadan2013, Chen2014}, our method relies on the {\it statistical properties} of cycles, as captured by Betti curves and persistence lifetime distributions, and is used as a generic tool for matrix analysis, rather than the analysis of point cloud data.  

We found that geometric organization of hippocampal place cell activity -- a prerequisite for the existence of spatial receptive fields -- can be detected from pairwise correlations alone, without any {\it a priori} knowledge about the nature of receptive fields.  Using simulated data from a model, we confirmed that such geometric structure would be observed as a result of realistic place fields, but would not arise from non-geometric (``scrambled'') place fields.  Perhaps surprisingly, we also found geometric organization in correlations during wheel running and REM sleep.  We suggest that clique topology is a powerful new tool for matrix analysis, and one that is especially useful in biological settings, to detect relevant structure in the presence of unknown nonlinearities.

\section*{Methods}
\paragraph{Order complex.}  For any $N \times N$ symmetric matrix $A$ with distinct entries, the order complex $\ord(A)$ is a sequence of graphs $G_0 \subset G_1 \subset \cdots \subset G_p$, where $G_0$ is the graph
having $N$ vertices and no edges, $G_1$ has a single edge $(ij)$ corresponding to the highest off-diagonal matrix value $A_{ij}$, and each subsequent graph has an additional edge for the next-highest off-diagonal matrix entry.  The graphs $\{G_k\}$ can also be indexed by the edge density,  $\rho = k/{{N \choose 2}}\in [0,1],$ 
where $k$ is the number of edges in the graph $G_k$. 

\paragraph{Betti curves.}  A clique in a graph $G$ is an all-to-all connected set of vertices in $G$. 
For each graph $G$ in the order complex $\ord(A)$, we compute simplicial homology groups $H_m(X(G),\mathbb Z_2)$ for $m = 1,2$ and $3$, where $X(G)$ is the clique complex of the graph $G$.
 We refer to this topological information as the {\it clique topology} of $G$, to distinguish it from the usual graph topology.  The dimensions  of the homology groups yield the Betti numbers, $\beta_m = \operatorname{dim} H_m(X(G),  \mathbb Z_2)$.  Indexing the graphs by edge density $\rho$,
 we organize  the Betti numbers across all graphs in the order complex into {\it Betti curves} $\beta_1(\rho), \beta_2(\rho),$ and $\beta_3(\rho)$.  The Betti curves provide a summary of the topological features of the matrix $A$.

\paragraph{Random and geometric matrices.}
{\it Random/shuffled} matrices were created by randomly permuting the ${N \choose 2}$ off-diagonal elements of $C$.  This is equivalent to considering random symmetric matrices with i.i.d. entries, whose corresponding order complex is a sequence of nested Erdos-Renyi random graphs. 
{\it Geometric} matrices  were obtained by sampling a set of $N$ i.i.d. points uniformly distributed in  the $d$-dimensional unit cube $[0,1]^d \subset \mathbb{R}^d,$ for $d \leq N$.  The matrix entries were then given by $C_{ij} =  - ||p_i - p_j||$, where the minus sign ensures that they monotonically {\it decrease} with distance, as expected for geometrically organized correlations.

\paragraph{Computation of pairwise correlation matrices.}  Given a set of simultaneously recorded spike trains from $N$ neurons, we computed cross-correlograms $\operatorname{ccg}_{ij}(\tau)$ for each pair of neurons $i,j$.  These functions were then normalized by the average firing rates of the neurons, and integrated over a timescale $\tau_{\max}$ in order to obtain an $N \times N$ matrix of pairwise correlations $C_{ij}$.  See Supplementary Methods and Supplementary Figure 3 for more details.

\section*{Acknowledgments} 
This work was supported by NSF DMS 1122519 (V.I.), 
 NSF DMS 1225666 (C.C.), a Sloan Research Fellowship (C.C.), and  
 the Howard Hughes Medical Institute (E.P.).  The authors also gratefully acknowledge 
 the Institute for Mathematics and its Applications, as well as 
 the Holland Computing Center at the University of Nebraska-Lincoln.

\bibliography{TDA_paper_references}

\begin{thebibliography}{10}

\bibitem{OKeefe1971}
J.~O'Keefe, J.~Dostrovsky, {\it Brain research\/} {\bf 34}, 171 (1971).

\bibitem{Hampson1996}
R.~E. Hampson, D.~R. Byrd, J.~K. Konstantopoulos, T.~Bunn, S.~A. Deadwyler,
  {\it Hippocampus\/} {\bf 6}, 281 (1996).

\bibitem{Caron2013}
S.~J.~C. Caron, V.~Ruta, L.~F. Abbott, R.~Axel, {\it Nature\/} {\bf 497}, 113
  (2013).

\bibitem{calcium-imaging2012}
C.~Grienberger, A.~Konnerth, {\it Neuron\/}  (2012).

\bibitem{Wigner1958}
E.~P. Wigner, {\it Ann. of Math. (2)\/} {\bf 67}, 325 (1958).

\bibitem{PasturShcherbina2011}
L.~Pastur, M.~Shcherbina, {\it Eigenvalue distribution of large random
  matrices\/}, vol. 171 of {\it Mathematical Surveys and Monographs\/}
  (American Mathematical Society, Providence, RI, 2011).

\bibitem{Hatcher}
A.~Hatcher, {\it Algebraic topology\/} (Cambridge University Press, Cambridge,
  2002).

\bibitem{EdelsbrunnerHarer}
H.~Edelsbrunner, J.~Harer, {\it Surveys on discrete and computational
  geometry\/} (Amer. Math. Soc., Providence, RI, 2008), vol. 453 of {\it
  Contemp. Math.\/}, pp. 257--282.

\bibitem{Kahle2009}
M.~Kahle, {\it Discrete Math.\/} {\bf 309}, 1658 (2009).

\bibitem{KahleMeckes2013}
M.~Kahle, E.~Meckes, {\it Homology Homotopy Appl.\/} {\bf 15}, 343 (2013).

\bibitem{ZomorodianCarlsson2005}
A.~Zomorodian, G.~Carlsson, {\it Discrete Comput. Geom.\/} {\bf 33}, 249
  (2005).

\bibitem{HarkerMischaikowMrozekNanda2014}
S.~Harker, K.~Mischaikow, M.~Mrozek, V.~Nanda, {\it Found. Comput. Math.\/}
  {\bf 14}, 151 (2014).

\bibitem{HUBELWIESEL1959}
D.~H. Hubel, T.~N. Wiesel, {\it The Journal of physiology\/} {\bf 148}, 574
  (1959).

\bibitem{Kahle2011}
M.~Kahle, {\it Discrete Comput. Geom.\/} {\bf 45}, 553 (2011).

\bibitem{gap}
C.~Curto, V.~Itskov, {\it PLoS Comput Biol\/} {\bf 4} (2008).

\bibitem{SinghMemoliIshkhanovSapiroCarlssonRingach2008}
G.~Singh, {\it et~al.\/}, {\it Journal of vision\/} {\bf 8}, 11 (2008).

\bibitem{NicolauLevineCarlsson2011}
M.~Nicolau, A.~J. Levine, G.~Carlsson, {\it Proceedings of the National Academy
  of Sciences\/} {\bf 108}, 7265 (2011).

\bibitem{DabaghianMemoliFrankCarlsson2012}
Y.~Dabaghian, F.~Memoli, L.~Frank, G.~Carlsson, {\it PLoS Comput Biol\/} {\bf
  8} (2012).

\bibitem{ChanCarlssonRabadan2013}
J.~M. Chan, G.~Carlsson, R.~Rabadan, {\it Proceedings of the National Academy
  of Sciences\/}  (2013).

\bibitem{Chen2014}
Z.~Chen, S.~N. Gomperts, J.~Yamamoto, M.~A. Wilson, {\it Neural computation\/}
  {\bf 26}, 1 (2014).

\bibitem{Pastalkova08}
E.~Pastalkova, V.~Itskov, A.~Amarasingham, G.~Buzs{\'a}ki, {\it {Science}\/}
  {\bf 321}, 1322 ({2008}).

\bibitem{WangLustigLeonardoRomaniPastalkova2014}
Y.~Wang, S.~Romani, B.~Lustig, A.~Leonardo, E.~Pastalkova, {\it Nature
  Neuroscience\/}  (2015).

\bibitem{JN11}
V.~Itskov, C.~Curto, E.~Pastalkova, G.~Buzsaki, {\it {Journal of
  Neuroscience}\/} {\bf 31} ({2011}).

\bibitem{GrosmarkMizusekiPastalkovaDibaBuzsaki2012}
A.~D. Grosmark, K.~Mizuseki, E.~Pastalkova, K.~Diba, G.~Buzsaki, {\it Neuron\/}
  {\bf 75}, 1001  (2012).

\bibitem{HillarWibisono2013}
C.~Hillar, A.~Wibisono, Maximum entropy distributions on graphs,
  arxiv:1301.3321 {[math.ST]}, \url{http://arxiv.org/abs/1301.3321} (2013).

\bibitem{Chazal:2013}
F.~Chazal, L.~J. Guibas, S.~Y. Oudot, P.~Skraba, {\it J. ACM\/} {\bf 60}, 41:1
  (2013).

\bibitem{Perseus}
V.~Nanda, The perseus software project for rapid computation of persistent
  homology, \url{http://www.sas.upenn.edu/~vnanda/perseus/index.html} (2013).

\bibitem{MischaikowNanda2013}
K.~Mischaikow, V.~Nanda, {\it Discrete Comput. Geom.\/} {\bf 50}, 330 (2013).

\bibitem{cliquetop}
C.~Giusti, Cliquetop: Matlab package for clique topology of symmetric matrices,
  \url{http://github.com/nebneuron/clique-top} (2014).

\end{thebibliography}
\bibliographystyle{Science}

\pagebreak

\section*{Supplementary Methods}

\subsection*{Data and preprocessing}
\paragraph{Experimental data.}  Spike trains of neurons in area CA1 of rodent hippocampus were  recorded during three behavioral conditions: (i) spatial navigation, (ii) wheel running and (iii) REM sleep.  Experimental procedures have been previously described in \cite{Pastalkova08,WangLustigLeonardoRomaniPastalkova2014}.   Spatial navigation was in a familiar, two-dimensional, 1.5m $\times$ 1.5m square box environment. Wheel running was in the context of a delayed alternation task, as described in \cite{Pastalkova08,JN11}.  
Periods of REM sleep were detected from the local field potential using the ratio of total delta power (0.1-3 Hz) to total theta power (5-10 Hz) in the spectrogram of the EEG \cite{GrosmarkMizusekiPastalkovaDibaBuzsaki2012}.  Periods with delta/theta ratio less than 1 were considered REM sleep.  

\paragraph{Selection criteria for cells and recordings.} During each of the three behavioral conditions, only putative pyramidal cells  whose  average firing rates were in the 0.2-7 Hz range were used.  Putative interneurons, defined as having an average firing rate above 7 Hz over an entire recording session, were   excluded.  Recordings with at least $N=60$ neurons satisfying these criteria were selected for the analyses.  
 A total of 18 recordings from 5 animals met the selection criteria.  These consisted of  9 ``open field'' spatial navigation data sets from three animals, 5 wheel running data sets from two animals, and 4 REM sleep data sets from two animals.  The number of cells and length of each recording were: N = 76 (16 min), N=81 (16 min), N=72 (63 min), N=88 (63 min), N=68 (59 min), N=64 (50 min), N=67 (50 min), N=74 (59 min), and N=66 (60 min) for spatial navigation; N=77 (26 runs, 6 min), N=73 (25 runs, 5 min), N=83 (13 runs, 5 min), N=67 (7 runs, 2 min), and N=64 (10 runs, 3 min) for wheel running; and
N=67 (30 s), N=69 (4 min), N=60 (2.5 min), and N=61 (3.5 min) for REM sleep.  The N=88 spatial navigation data set was used in Figure 3.  

\paragraph{Computation of the pairwise correlation matrices.} 
The cross-correlograms  were computed  as $\operatorname{ccg}_{ij}(\tau)=\frac1{T} \int_{0}^{T}  f_i(t)f_j(t+\tau) \text{d}t$, where $f_i(t)$   is  the firing rate of the $i$-th neuron and $T$ is the total duration of the considered time period. 
The normalized cross-correlation on a timescale of $\tau_{\operatorname{max}} $   was computed  as 
$$  C_{ij}=\frac 1 {\tau_{\operatorname{max}} {r_i r_j}}\max\left( \int _0 ^ {\tau_{\operatorname{max}}} \! \! \! \! \! \!  \operatorname{ccg}_{ij}(\tau) d \tau, \int _0 ^ {\tau_{\operatorname{max}}} \! \! \! \! \! \!  \operatorname{ccg}_{ji}(\tau) d \tau\right),  
 $$ 
 where $r_i$ is the average firing rate of the $i$-th neuron (see Supplementary Figure 3).

  \paragraph{Simulated spike train data from PF and scrambled PF models.} 
 
For each cell in the N=88 spatial navigation data set, place fields $F_i({\bf x})$ were computed using a 100 $\times$ 100 grid of pixels.    In each pixel, the number of spike events for each cell was normalized by the time spent at that location, and then smoothed with a 2-dimensional Gaussian ($\sigma$ = 5 grid locations, see also Supplementary Figure 9).  Simulated spike trains for the PF model were generated from the place fields as inhomogeneous Poisson processes with rate functions $r_i(t) = F_i({\bf x}(t))$, using the animal's original trajectory, ${\bf x}(t)$.

Scrambled place fields $\widetilde{F}_i({\bf x})$ were computed by randomly permuting the pixels in the 100 $\times$ 100 grid.  A different permutation was used for each place field, in order to destroy the coherence of the spatial organization across the population.  Similar spike trains for the scrambled PF model were again generated as inhomogeneous Poisson processes using the original trajectory, ${\bf x}(t)$, but with rate functions $r_i(t) = \widetilde{F}_i({\bf x}(t))$ given by the scrambled place fields (see Supplementary Figure 9). 
The simulated data sets were analyzed in Figure 3 at the correlation timescale $\tau_{\max}=1$sec.

\subsection*{Random and geometric matrices}

For each symmetric matrix $C$ we consider three types of controls: shuffled (or ``random'') control matrices, weighted-maximum-entropy (WME) control matrices, and geometric matrices. 

{\bf Shuffled matrices}  were created by randomly permuting the ${N \choose 2}$ off-diagonal elements of $C$.  Because only the ordering of matrix elements is considered in the subsequent topological analyses, this is equivalent to considering random symmetric matrices with i.i.d. entries, whose corresponding order complex is a sequence of nested Erdos-Renyi random graphs. 

{\bf WME matrices}  were  obtained by sampling the maximum entropy distribution on weighted graphs with constrained mean degree sequence induced by $C$. This distribution was previously described in \cite{HillarWibisono2013} and is estimated by summing the rows of $C$. The distribution on each element $(i,j)$ in the matrix  is exponential with mean $\frac{1}{\theta_i + \theta_j}$. The parameters $\theta_i$ were  obtained by 
 solving the system of equations 
 $$
\sum_{j\neq i} \frac1{\theta_i+\theta_j}=\sum_{j\neq i} C_{ij}, \quad \text{ for } i=1,\dots, N,
 $$
using  standard gradient descent methods \cite{HillarWibisono2013}.   

{\bf Geometric matrices}   were obtained by sampling a set of $N$ i.i.d. points uniformly distributed in  the $d$-dimensional unit cube $[0,1]^d \subset \mathbb{R}^d,$ for $d \leq N$.  The matrix entries were then given by $C_{ij} =  - ||p_i - p_j||$, where the minus sign ensures that they monotonically {\it decrease} with distance, as expected for geometrically organized correlations.

\subsection*{Clique topology}
We performed topological data analysis on pairwise correlation matrices.  Here we describe the general  procedure;  for more detailed explanations, see the Supplementary Text.  Our approach relies on the statistical properties of homology cycles,  as captured by Betti curves, in order to detect structure (or randomness) in symmetric matrices.  Note that even for data reflecting a two-dimensional space, the higher-dimensional Betti curves $\beta_2(\rho)$ and $\beta_3(\rho)$ were crucial for 
detecting the presence or absence of geometric organization (Figure 3), while $\beta_1(\rho)$ was less informative.\footnote{The Betti curve $\beta_0(\rho)$, which we have not used here, counts the number of connected components in each clique complex, and may thus be useful for clustering \cite{Chazal:2013}.}

\paragraph{Order complex.}  For any $N \times N$ symmetric matrix $A$ with distinct entries, the order complex $\ord(A)$ is a sequence of graphs $G_0 \subset G_1 \subset \cdots \subset G_p$, where $G_0$ is the graph
having $N$ vertices and no edges, $G_1$ has a single edge $(ij)$ corresponding to the highest off-diagonal matrix value $A_{ij}$, and each subsequent graph has an additional edge for the next-highest off-diagonal matrix entry.  The graphs $\{G_k\}$ can also be indexed by the edge density, $\rho = k/{{N \choose 2}}\in [0,1],$ 
where $k$ is the number of edges in the graph $G_k$.

\paragraph{Betti curves.} 

A clique in a graph $G$ is an all-to-all connected set of vertices in $G$. 
For each graph $G$ in the order complex $\ord(A)$, we compute simplicial homology groups $H_m(X(G),\mathbb Z_2)$ for $m = 1,2$ and $3$, where $X(G)$ is the clique complex of the graph $G$.
 We call this the {\it clique topology} of $G$, to distinguish it from the usual graph topology.  The dimensions  of the homology groups $H_m(X(G),  \mathbb Z_2)$,   yield the Betti numbers $\beta_m$.  Indexing the graphs by edge density $\rho$,
 we organize  the Betti numbers across all graphs in the order complex into {\it Betti curves} $\beta_1(\rho), \beta_2(\rho),$ and $\beta_3(\rho)$.  The Betti curves provide a summary of the topological features of the matrix $A$.

To compute Betti curves for a matrix $A$, we begin by finding all maximal cliques of up to five vertices (those are needed to compute $\beta_3$) for  each graph $G_\rho$,  with $\rho \leq 0.6$. The resulting lists are then input into Perseus, a computational topology software package implemented by Vidit Nanda \cite{Perseus}; this software builds on  work by Mishaikow and Nanda \cite{MischaikowNanda2013} using  discrete Morse theory to reduce the sizes of simplicial complexes before performing persistent homology computations.
All software used in this process  is available in the Matlab package CliqueTop \cite{cliquetop}.

\paragraph{Integrated Betti values.}  In order to facilitate the comparison of Betti curves to control matrices, we integrate the Betti curves with respect to graph density:
$$ \bar{\beta}_m= \int_{0}^1 \beta_m(\rho) \;\text{d}\rho.$$
The values $\bar{\beta}_1$, $\bar{\beta}_2$, and $\bar{\beta}_3$ were computed for each data set.  For distributions of shuffled and geometric control Betti curves, the resulting integrated Betti values are summarized in box-and-whisker plots.  We used standard box plots in Matlab, with bottom, middle, and top horizontal lines on the boxes denoting first quartile ($Q_1$, 25\%-ile), median (50 \%-ile), and third quartile ($Q_3$, 75\%-ile) boundaries in the distributions of integrated Betti values; while the bottom and the whiskers denote $Q_1-1.5(Q_3-Q_1)$ and $Q_3+1.5(Q_3-Q_1)$ respectively.  

\paragraph{Significance threshold.}  Our threshold for rejecting the geometric hypothesis for a given integrated Betti value was obtained from the box-and-whisker plot for a distribution of 100 geometric matrices with matching $N$ and dimension $d=N$.  Specifically, we used the top whisker value, $Q_3+1.5(Q_3-Q_1)$, as the significance threshold.  The bottom whisker was not used, as Betti values lower than this are consistent with geometric matrices with smaller dimension $d$.  In a normal distribution, 99.3 \% of the data lie within the whiskers, so that less than 0.4 \% of data points lie above the top whisker.  Our integrated Betti values $ \bar{\beta}_m= \int_{0}^1 \! \!  \beta_m(\rho) \text{d}\rho$ for geometric controls, however, are not normally-distributed.  In the case of $\bar{\beta}_1$ and $\bar{\beta}_2$, the top whisker corresponds, on average, to the 98 \%-ile of the distribution.  In the case of $\bar{\beta}_3$, the top whisker is just under the 97 \%-ile value.  A data point above the top whisker is thus inconsistent with geometric controls with $p < 0.05$. For comparisons against shuffled/random control distributions, we computed the $p$-value directly from the distribution, as in these cases we built the distributions from 1000 trials, rather than just 100.  Note that clique topology computations are much faster for matrices with random structure than for geometric matrices, because of differences in the statistics of the cliques.
 
 \pagebreak

\section*{Supplementary Figures}
\bigskip

\begin{figure*}[!h]
\begin{center}
\includegraphics[width=4in]{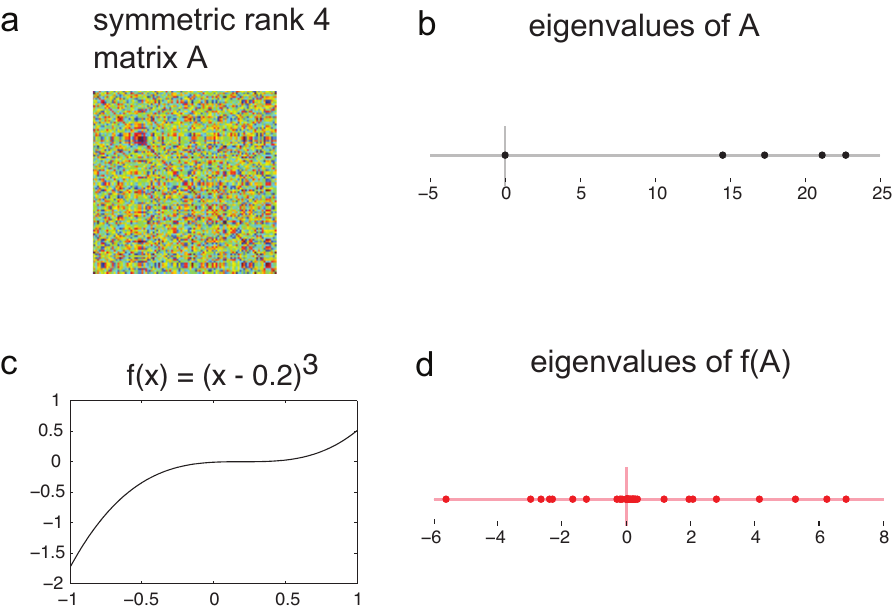}
\end{center}
 
 {\bf Supplementary Figure 1: Spectral signatures of matrix structure   are destroyed by nonlinear monotonically increasing transformations.} 
 (\textbf{a}) A  $100 \times 100$ symmetric matrix $A$ with $\operatorname{rank}(A) =4$. 
(\textbf{b})   The spectrum of $A$ includes four nonzero eigenvalues that are all positive.  This is the signature that $A$ has rank 4 and is positive semidefinite.
(\textbf{c}) The graph of the monotonically increasing function $f(x)=(x-0.2)^3$.  
(\textbf{d}) The spectrum of the matrix $f(A)$ contains many nonzero eigenvalues.  The spectral signature that $A$ has low-rank structure has been destroyed by $f$.
\end{figure*}

\pagebreak

\begin{figure}[!h]\label{sfig2}
\begin{center}
\includegraphics[width=6in]{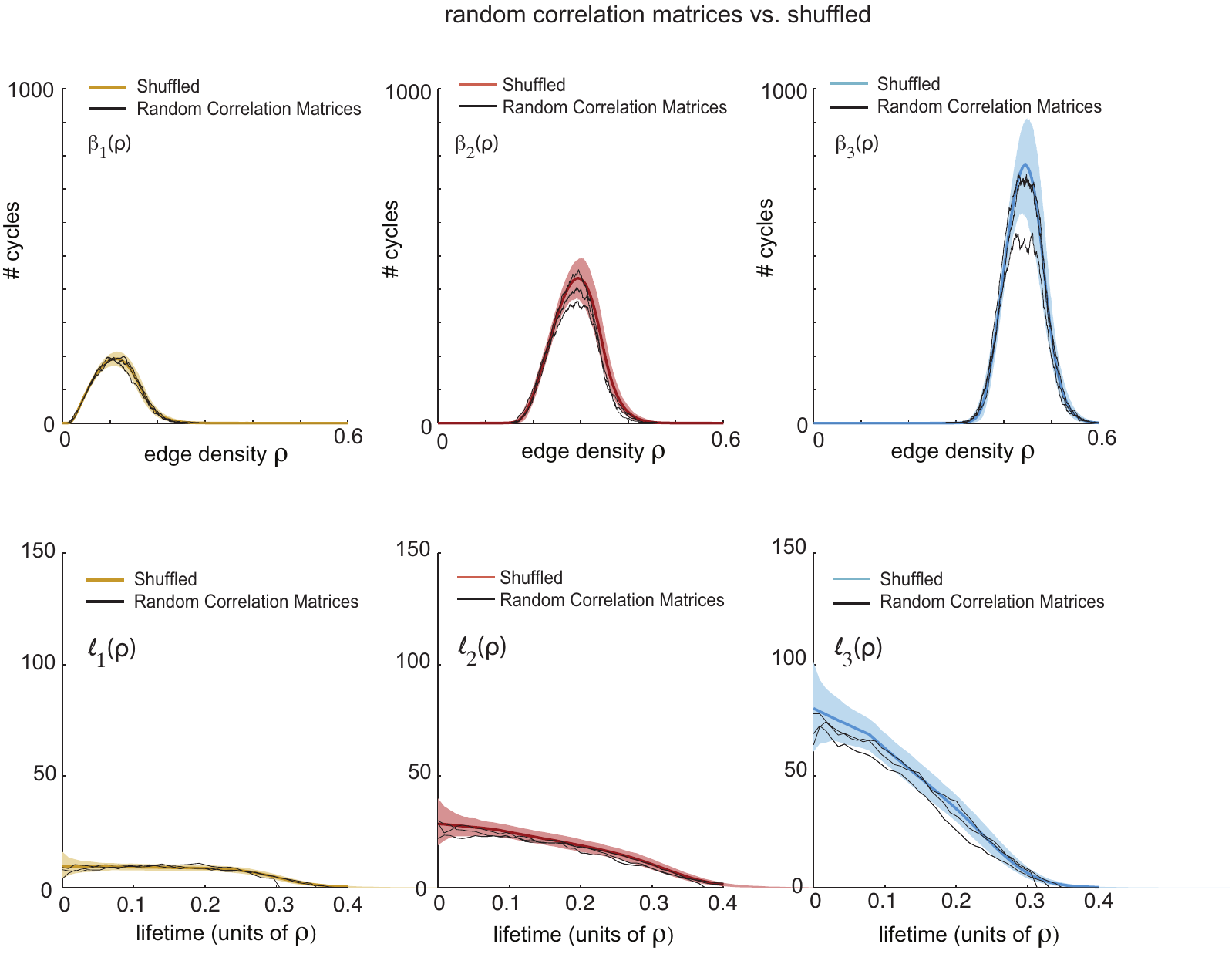} 
\end{center}

{\bf Supplementary Figure 2:}  { \bf Topological properties of random  correlation matrices are similar to those of random i.i.d.  matrices.} 
The $N\times N$  correlation matrix $C_{ij}=\operatorname{corr}(X_i,X_j)$ was computed using  $10,000$  samples of  $N=88$  independent uniformly distributed random variables $X_i$.
Each panel compares  the  Betti curves (top) and the persistence lifetimes (bottom) of the random correlation matrices to those of the random (shuffled) matrices. Each of the three black lines   correspond to one instance of such a  correlation matrix. Colored lines and shaded regions  correspond to the mean curves and 95\% confidence intervals for the random (shuffled) matrices. 
\end{figure}

\pagebreak

\begin{figure}[!h]\label{sfig3}
\begin{center}
\includegraphics[width=4in]{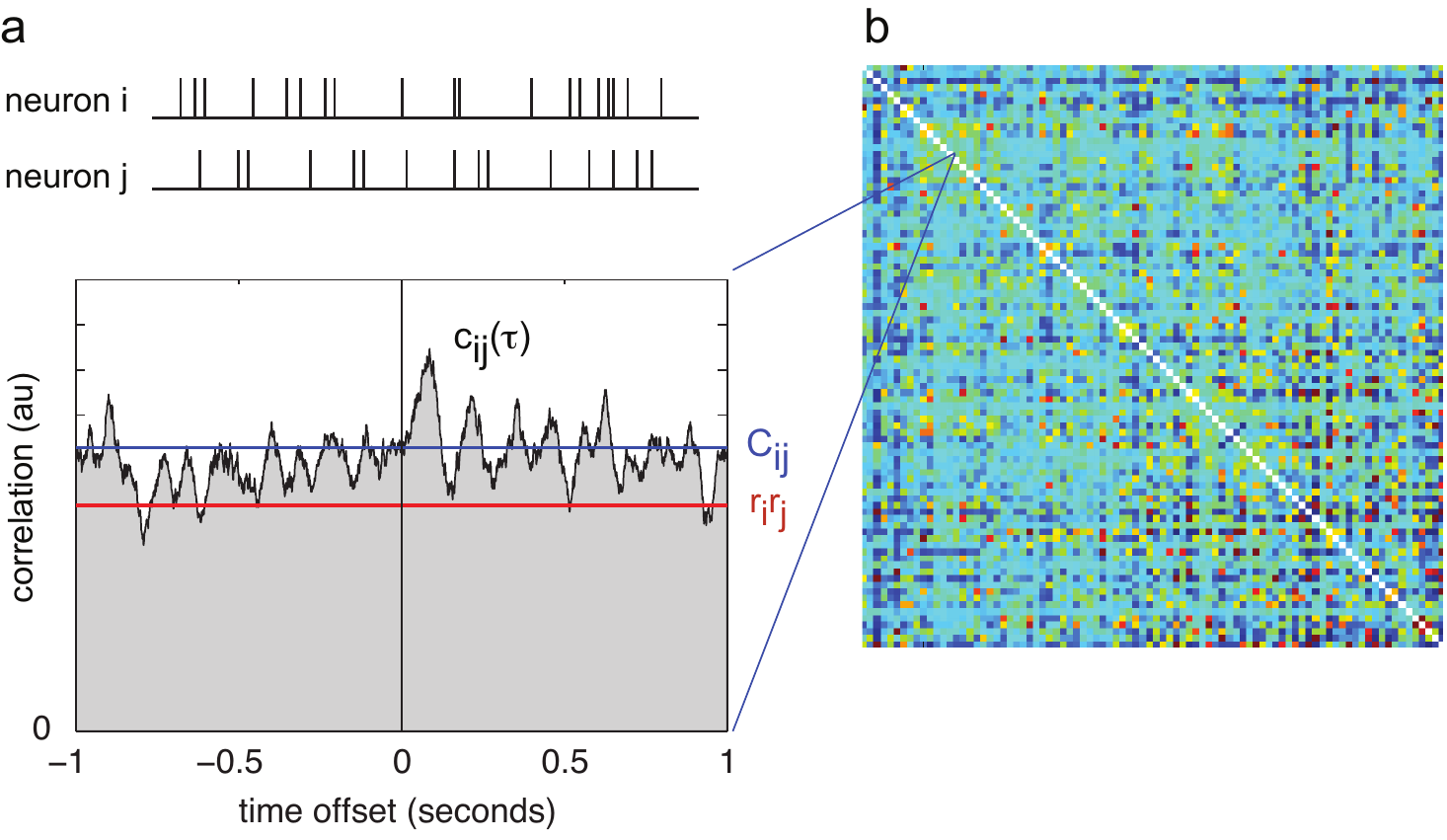}
\end{center}

\bigskip 
 {\bf Supplementary Figure 3:} {\bf Computation of pairwise correlation matrices from spike train data.} \textbf{(a)} For a pair of spike trains 
 $ \{ t^i_\ell\}_{\ell=1\dots n_i}$ and    $ \{ t^j_\ell\}_{\ell=1\dots n_j}$ for neurons $i$ and $j$ (top),  the cross-correlogram $\operatorname{ccg}_{ij}(\tau)$  is computed as 
 $$\operatorname{ccg}_{ij}(\tau)=\frac1{T} \int_{0}^{T}  f_i(t)f_j(t+\tau) \text{d}t,$$  
  where 
  $f_i(t)=\sum_{\ell=1}^{n_i} \delta(t-t^{i}_\ell)$ is the instantaneous firing rate of the $i$-th neuron. The graph of a smoothed $ \operatorname{ccg}_{ij}(\tau)$ is displayed (black curve) along with the expected value of the cross-correlogram, $r_i r_j$ (red), for uncorrelated spike trains with matching firing rates,  $r_i=n_i/T$.
    The pairwise correlations $C_{ij}$ (blue line), with timescale $\tau_{\max}$, were computed as 
  $$  C_{ij}=\frac 1 {\tau_{\operatorname{max}} {r_i r_j}}\max\left( \int _0 ^ {\tau_{\operatorname{max}}} \! \! \! \! \! \!  \operatorname{ccg}_{ij}(\tau) d \tau, \int _0 ^ {\tau_{\operatorname{max}}} \! \! \! \! \! \!  \operatorname{ccg}_{ji}(\tau) d \tau\, \right) 
 $$   
 The timescale $\tau_{\max} = 1$ sec was used in all but one panel of Figures 3 and 4, while a range of timescales from 10 ms to 2 sec appears in Figure 3d.
 \textbf{(b)}   The $88 \times 88$ matrix $C$ for the spatial exploration data used in  Figure 3a,b,c.  The entry $C_{14,15}$ corresponds to the cross-correlogram in panel (a).
  
  \end{figure}

\pagebreak

\begin{figure}[!h] 
\begin{center}
\includegraphics[width=6in]{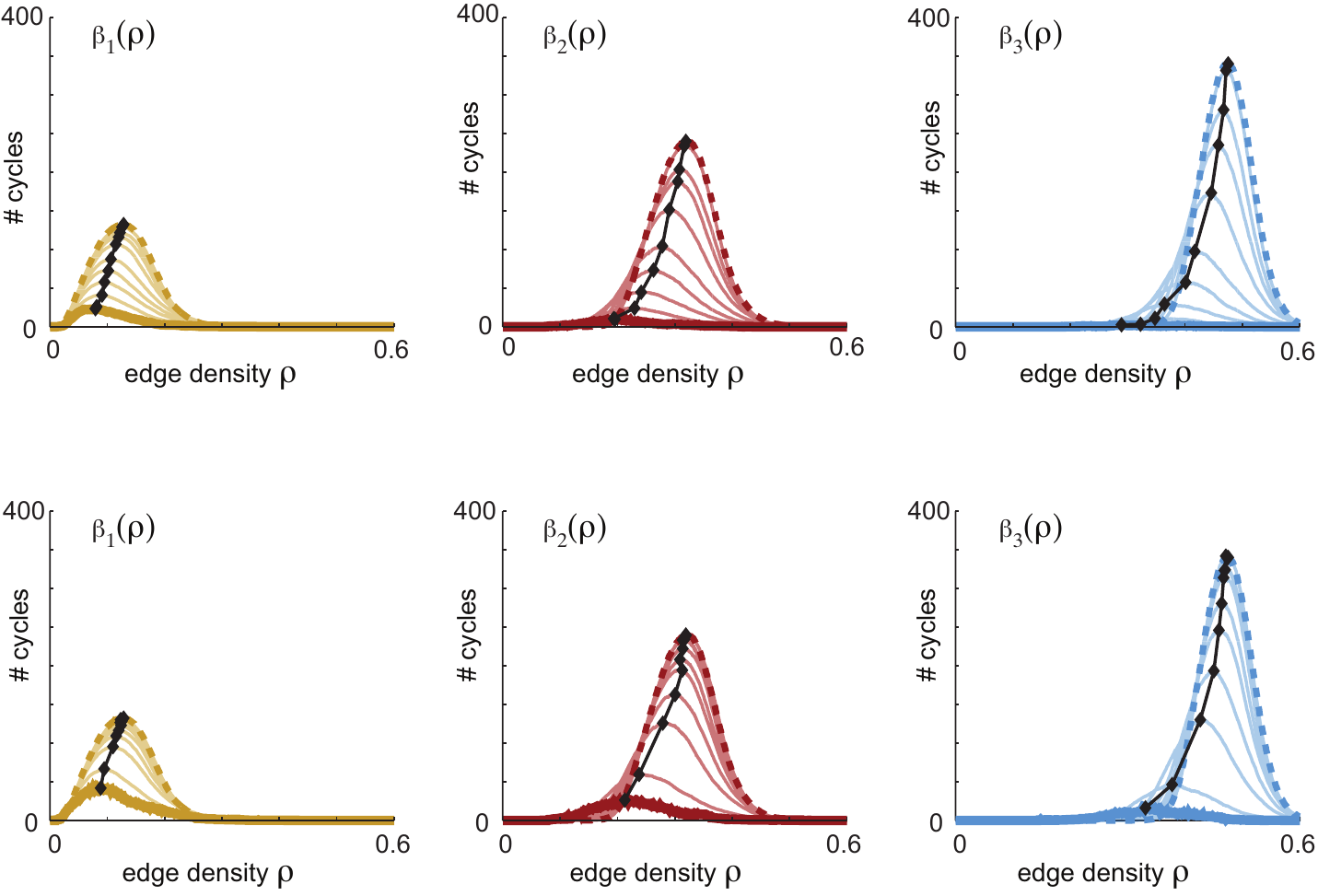}
\end{center}
\bigskip 
 {\bf Supplementary Figure 4:} {\bf  The rightward shift of the Betti curves can be explained by adding noise to ``noiseless'' geometric matrices.}  $N=70$  random independent uniformly distributed points $p_i$ in a $d$-dimensional unit cube were sampled in dimensions $d=10$ and $d=N$.  Noised geometric matrices were obtained as 
 $$ A_{ij}=-\Vert p_i-p_j\Vert +(\nu\sqrt{d}) g_{ij},$$
 where the  terms $g_{ij}$ were independent and normally distributed with zero mean and unit variance, 
 and the normalized noise strength  $\nu$ took  the following values: $0.01, 0.02, 0.03, 0.04, 0.05, 0.1, 0.25 , 0.5$. Note that for $\nu = 0.5$ the matrix entries are dominated by the noise term, and $A_{ij}$ is close to a random matrix. 
Each panel compares  the  Betti curves ($d=10$ top, and $d=N$ bottom)  of the noised matrices (thin curves, stratified by the magnitude of  the parameter $\nu$)   to those of the random matrices (dashed lines) and also the zero noise geometric matrices (thick curves).  Black lines  connect the maximum values of the mean curves for the different levels of noise. Small amounts of noise produce a rightward shift in the peak values of the Betti curves, while curves for $\nu = 0.5$ are indistinguishable from those of random matrices.
 \end{figure}
 
 \pagebreak

\begin{figure}[!h] 
\begin{center}
\includegraphics[width=4in]{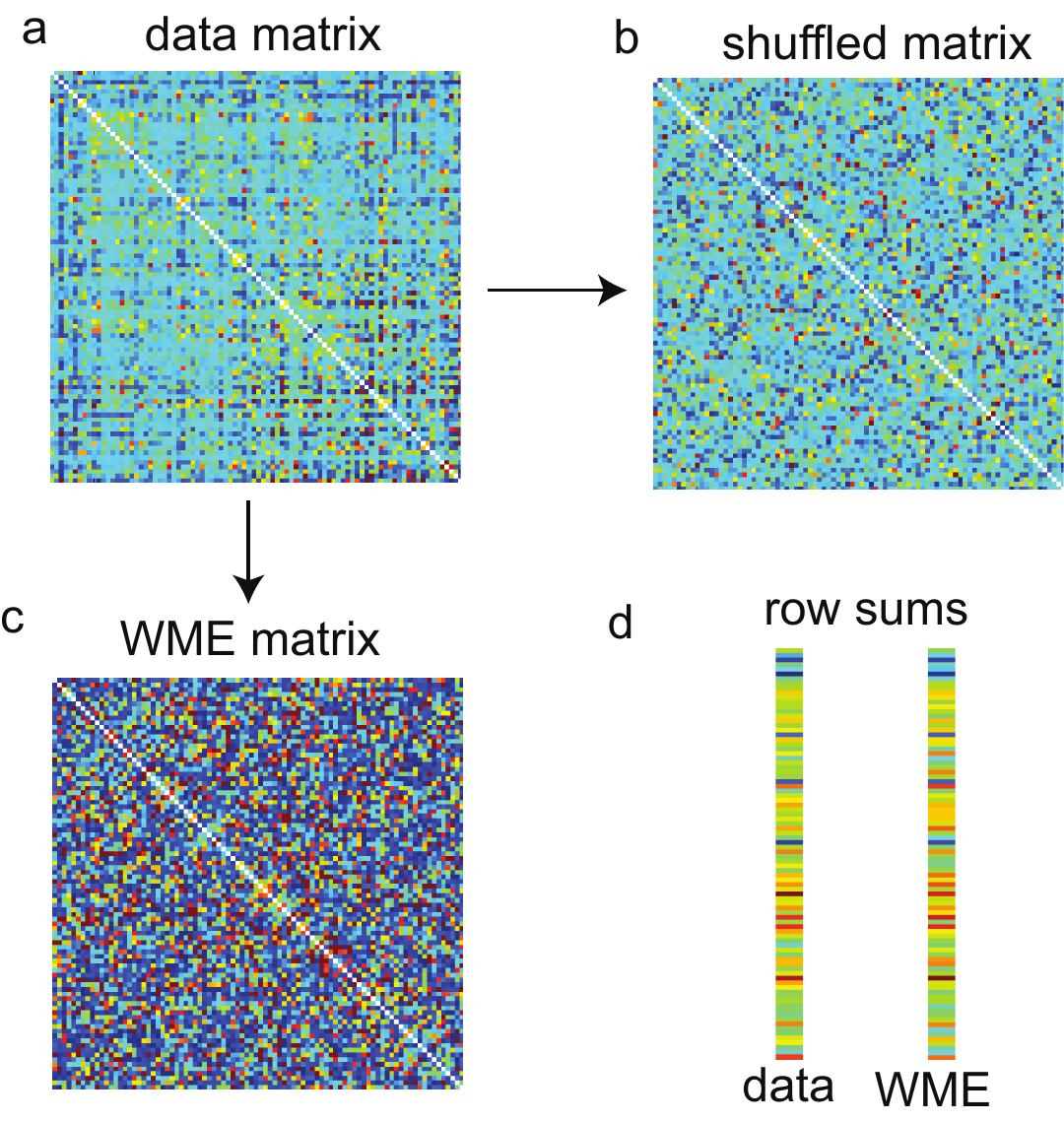}
 \end{center}

\bigskip 
 {\bf Supplementary Figure 5:} {\bf Two types of random controls.}  (\textbf{a}) The matrix $C_{ij}$ used in Figure 3a. 
 (\textbf{b}) A shuffled matrix obtained by randomly permuting the ${88 \choose 2}$ off-diagonal elements of the symmetric matrix in panel (a). 
 (\textbf{c}) A sample from the maximum entropy (WME) distribution on $88 \times 88$ symmetric matrices with  prescribed expected values of row sums matching those of the matrix in panel (a). The distribution on each element $(i,j)$ in the matrix  is exponential with mean $\frac{1}{\theta_i + \theta_j}$ (Hillar \&  Wibisono, 2013  \text{http://arxiv.org/abs/1301.3321}). The parameters $\theta_i$ were  obtained by 
 solving the system of equations 
 $$
\sum_{j\neq i} \frac1{\theta_i+\theta_j}=\sum_{j\neq i} C_{ij}, \quad \text{ for } i=1,\dots, N,
 $$
using  standard gradient descent methods (Hillar \&  Wibisono, 2013).   
 (\textbf{d}) (left) The row sums $\sum_{j\neq i} C_{ij} $ for the matrix in (a);  (right) mean row sums for twenty samples from the distribution described in (c).
\end{figure}

\pagebreak

\begin{figure}[!h] 
\begin{center}
\includegraphics[scale=1.4]{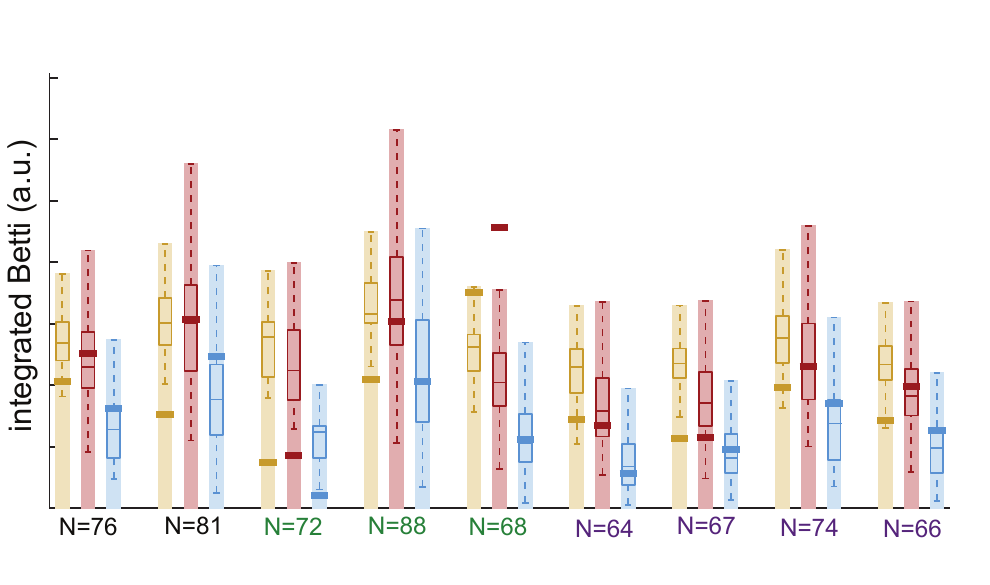}
\end{center}
 {\bf Supplementary   Figure 6:} {\bf   Integrated Betti values obtained from neural activity during spatial navigation  are  consistent with those of geometric matrices.}  
  Integrated Betti values from place cell data are compared to geometric distributions with matching $N$ across nine recordings of rat hippocampus during spatial navigation, obtained from three animals.  
 The geometric box plots are shown for the  dimension, $d = N$, while the shaded area indicates the confidence interval across the dimensions $d\leq N$.  
  The number $N$ of neurons is displayed in color (black, green, and purple) to indicate recordings from the same animal.  Betti values for the place cell data are  consistent with those of geometric matrices in all but one data set ($N = 68$).  
  \end{figure}

\pagebreak

\begin{figure}[!h] 
\begin{center}
\includegraphics[scale=1.6]{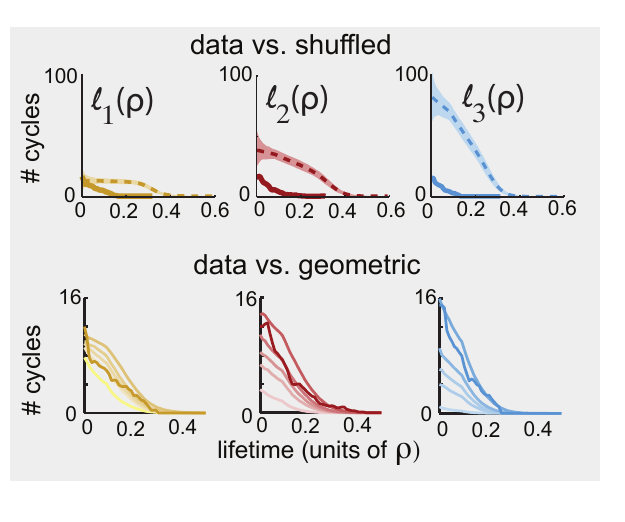}
\end{center}
\bigskip 
 {\bf Supplementary Figure 7:} {\bf  The persistence lifetime distributions, computed from neural activity during  spatial navigation  resemble the geometric controls but are significantly below those of the shuffled matrices.}  
Persistence lifetime distributions, $\ell_1(\rho)$ (yellow), $\ell_2(\rho)$ (red) and $\ell_3(\rho)$ (blue), computed from the same data set and controls as those used in Figure 3b.  
(Top)
The lifetime distributions for the data (solid lines) fall off quickly, while those of the shuffled matrices are much broader (dashed lines are means over 1000 trials; shading shows the 95\% confidence intervals).  
(Bottom) Here the data lifetime distributions (solid lines) are overlaid with the mean distributions (faint lines) for 1000 geometric matrices in each dimension $d =5, 10, 16, 24,$ and $88$.  As with the geometric Betti curves, the persistence lifetime distributions for geometric matrices are stratified by dimension, with the top curves corresponding to the highest dimension.  

 \end{figure}
  
\pagebreak
\begin{figure}[!h] 
\begin{center}
\includegraphics[width=6in]{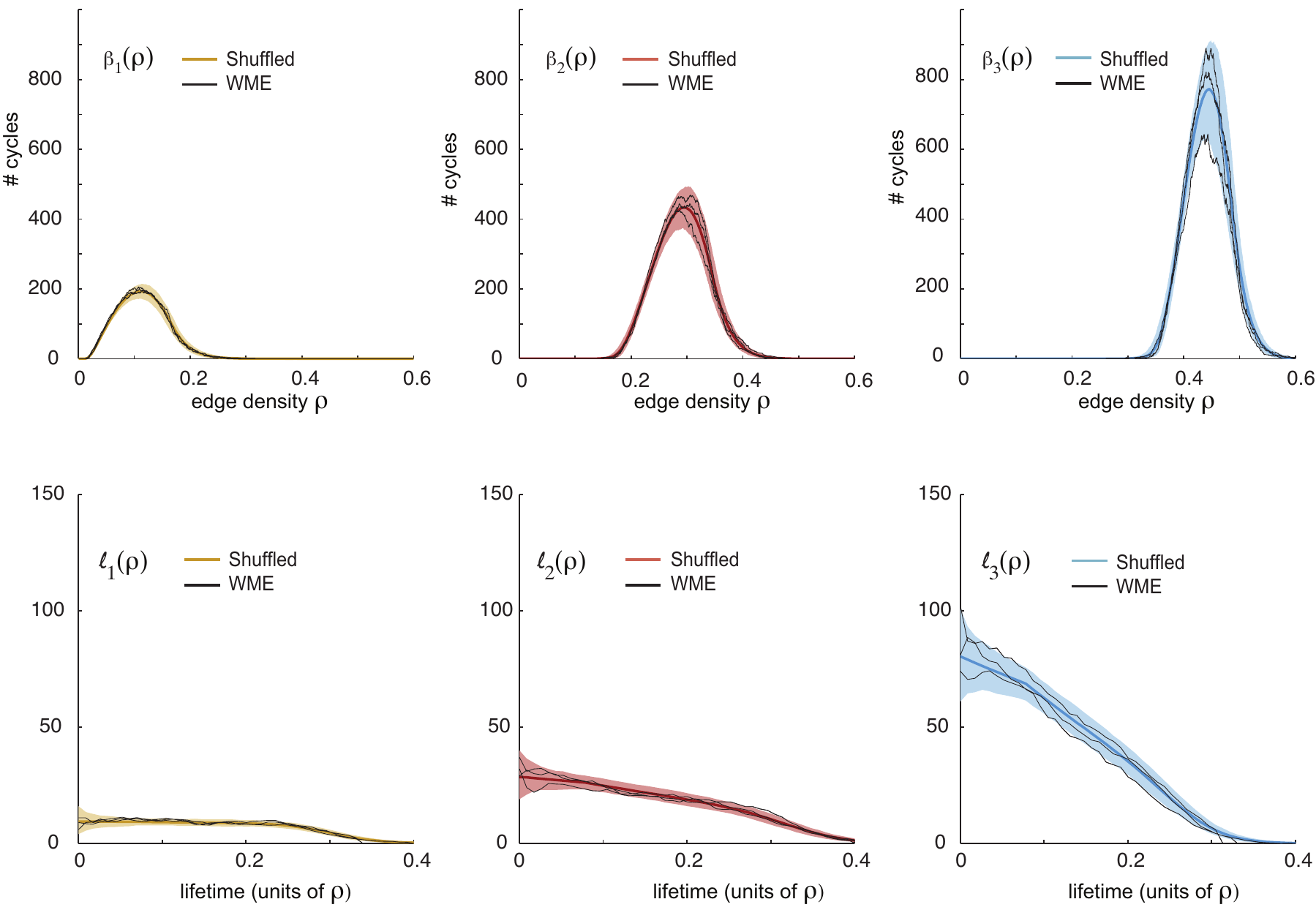}
 \end{center}
\bigskip 
 {\bf Supplementary Figure 8:} {\bf  Clique topology of  WME matrices, sampled from the maximum entropy distribution with prescribed row sums, is similar to that of random (shuffled) matrices.}  Comparison of Betti curves (top) and persistence lifetimes (bottom) for WME matrices computed using the matrix $C_{ij}$ used in Figure 3a.   Each panel compares  the  Betti curves (top) and the persistence lifetimes (bottom) of the WME  matrices,  computed using the matrix $C_{ij}$ used in Figure 3a, to those of the random (shuffled) obtained from the same $C_{ij}$. Each of the three black lines   correspond to one sampling of a WME   matrix. Colored lines and shaded regions  correspond to the mean curves and 95\% confidence interval for the  shuffled matrices.  \end{figure}
  
\pagebreak

\begin{figure}[!h]\label{sfig9}
\begin{center}
\includegraphics[width=4in]{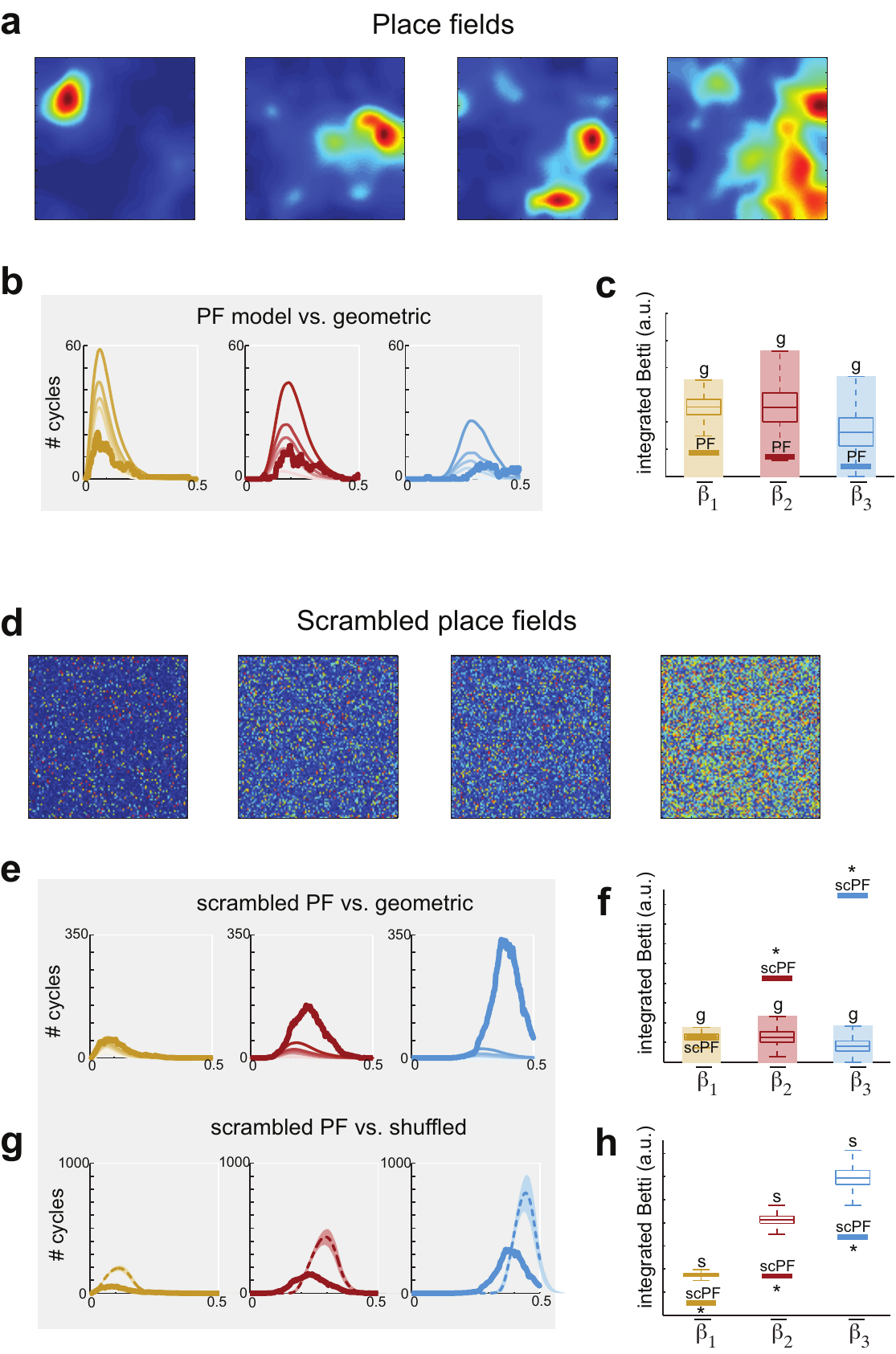}
\end{center}
\begin{small}
{\bf Supplementary Figure 9:}    (\textbf{a})  Example of several place fields computed from spike trains during spatial exploration in the N=88 data set. 
(\textbf{b}) Betti curves (bold lines) computed from the simulated spike trains ($\tau_{\max}=1sec$)  for the place field model versus the geometric Betti curves (thin lines stratified by dimensions). 
(\textbf{c}) Integrated Betti values (bold lines, labelled PF) for the curves in panel (b)  lie in the geometric regime, in agreement with those of the original data.  
(\textbf{d})  The scrambled place fields corresponding to those in panel (a). These were obtained by sub-dividing the square into a $100\times 100$ grid and randomly permuting each pixel. The permutations were independent for each cell.    
(\textbf{e})  Betti curves (bold lines) derived from the spike trains ($\tau_{\max}=1sec$)  generated using the scrambled PF model versus the geometric Betti curves (thin lines stratified by dimensions).  
(\textbf{f})  Integrated Betti values  from the scrambled PF model (bold lines, labelled scPF) lie outside of  the  the significance threshold (see Supplementary Methods)  for the geometric regime for   $\bar \beta_2$ and $\bar \beta_3$.
(\textbf{g})   Betti curves (bold lines)  derived from the spike trains ($\tau_{\max}=1sec$)  generated using the  Scrambled PF model are significantly smaller than those of shuffled controls.
(\textbf{h}) Integrated Betti values  derived from from the scrambled PF model (bold lines, labelled scPF)  
 are outside the 99.9\% confidence intervals for the shuffled matrices.  Box plots for shuffled matrices are the same as in Figure 3c.
\end{small}
\end{figure}

\pagebreak
\begin{figure}[!h]\label{sfig10}
\begin{center}
\includegraphics[width=6in]{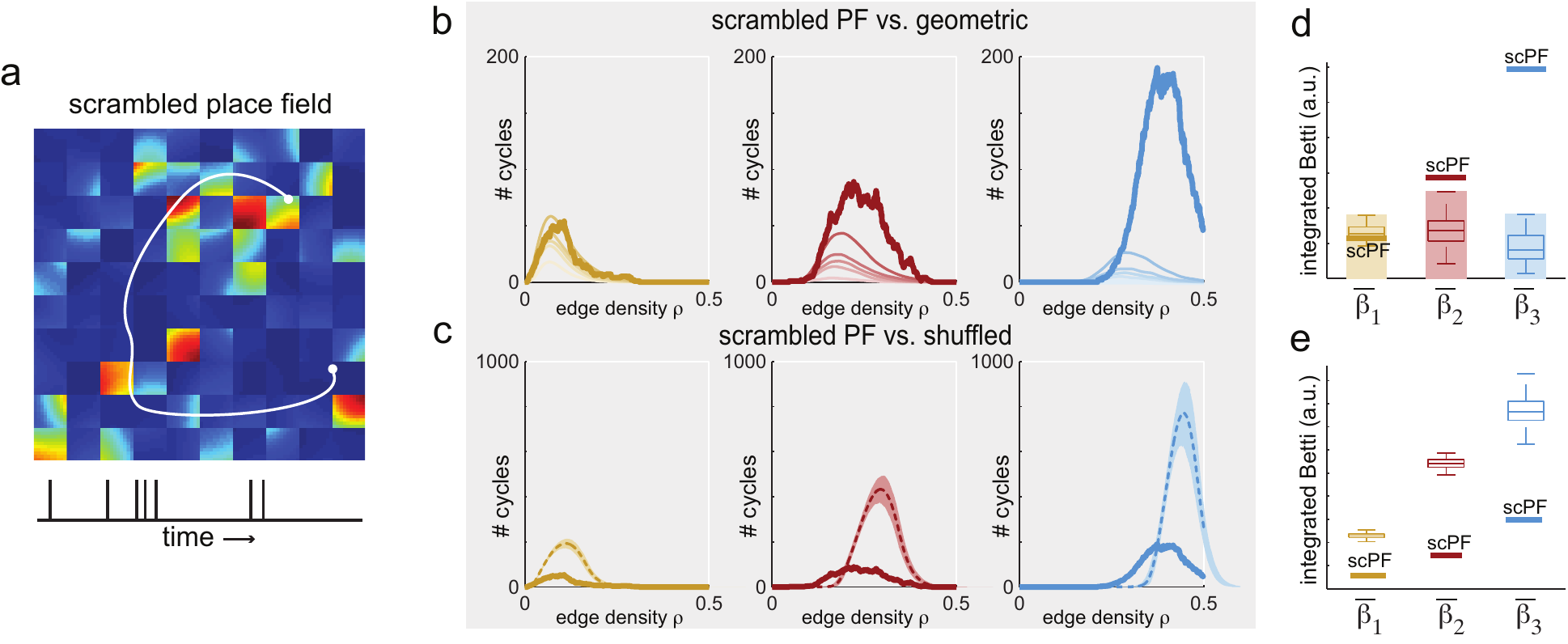} 
\end{center}
\bigskip
{\bf Supplementary Figure 10:} { \bf Betti curves and summary stats for the 10x10 grid  scrambled place fields.} 
(\textbf{a})  A single scrambled place field corresponding to the place field in Figure 3e. The scrambling is performed on a   $10\times 10$ grid. A cartoon trajectory (white) is displayed  together with the corresponding spike train (bottom).
(\textbf{b}) Betti curves from the scrambled place field  model (bold lines) lie outside the geometric regime for $\beta_2$ and $\beta_3$.
(\textbf{c}) Betti curves for the scrambled place field  model are   significantly smaller than the shuffled controls.
(\textbf{d}) Integrated Betti values (bold lines, labelled scPF) for the scrambled place field model also lie outside of  the significance threshold (see Supplementary Methods)  for the geometric regime for $\bar{\beta}_2$ and $\bar{\beta}_3$,
while $\bar{\beta}_1$ is in the geometric regime.
(\textbf{e}) Integrated Betti values (bold lines, labelled scPF) for the scrambled place field  model are outside the 99.9\% confidence intervals for the shuffled matrices.  Box plots for geometric and shuffled matrices are the same as in Figure 3c.
\end{figure}

\pagebreak

\begin{figure}[!h]\label{sfig11}
\begin{center}
\includegraphics[width=6in]{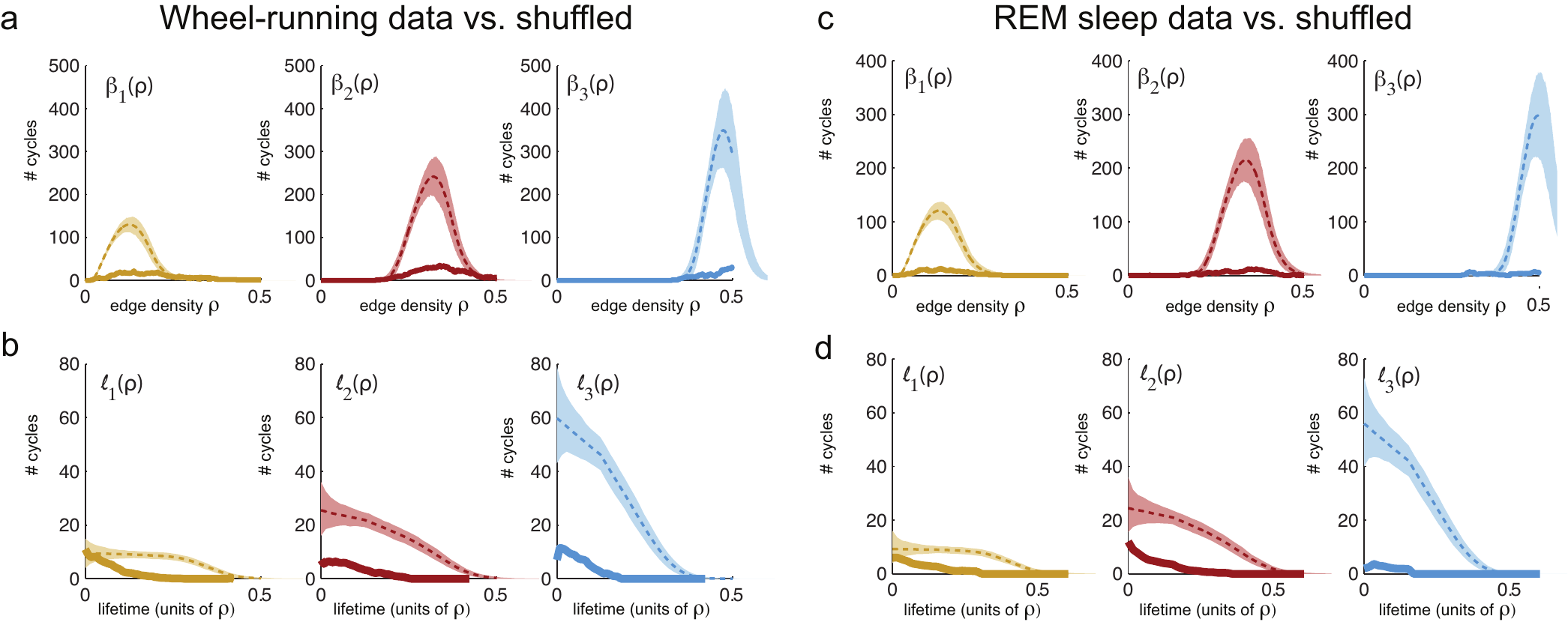}
 \end{center}
\bigskip 
 {\bf Supplementary Figure 11:}  {\bf The clique topology of spike train correlations during wheel running and REM sleep is significantly non-random as compared to shuffled matrices.}  
 (\textbf{a}) Comparison of the data Betti curves (solid lines) for the spike trains during wheel running  (N=73 in Figure 4a) to those of shuffled correlation matrices (dashed lines are means over 1000  trials and shading shows 95\% confidence intervals, as in Figure 1e).  For each $m=1,2,3$, the data Betti curves are orders of magnitude smaller than those of the shuffled curves.
(\textbf{b}) Persistence lifetime distributions (solid lines) for the same data as in (a), $\ell_1(\rho)$ (yellow), $\ell_2(\rho)$ (red) and $\ell_3(\rho)$ (blue).    The lifetime distributions for the data  fall off quickly, while those of the shuffled matrices are much broader (dashed lines are means over 1000 trials; shading shows the 95\% confidence intervals).  
(\textbf{c}) Comparison of the  Betti curves (solid lines) for spike train correlations during REM sleep (N= 67 in Figure 4b) to those of shuffled correlation matrices (dashed lines are means over 1000  trials and shading shows 95\% confidence intervals, as in Figure 1e).  For each $m=1,2,3$, the data Betti curves are orders of magnitude smaller than those of the shuffled curves.
(\textbf{d}) Persistence lifetime distributions (solid lines) for the same data as in (c).    The lifetime distributions for the data  fall off quickly, while those of the shuffled matrices are much broader (dashed lines are means over 1000 trials; shading shows the 95\% confidence intervals).  
 \end{figure}

\newpage
 \begin{figure}[!h]\label{sfig12}
\begin{center}
\includegraphics[width=6 in]{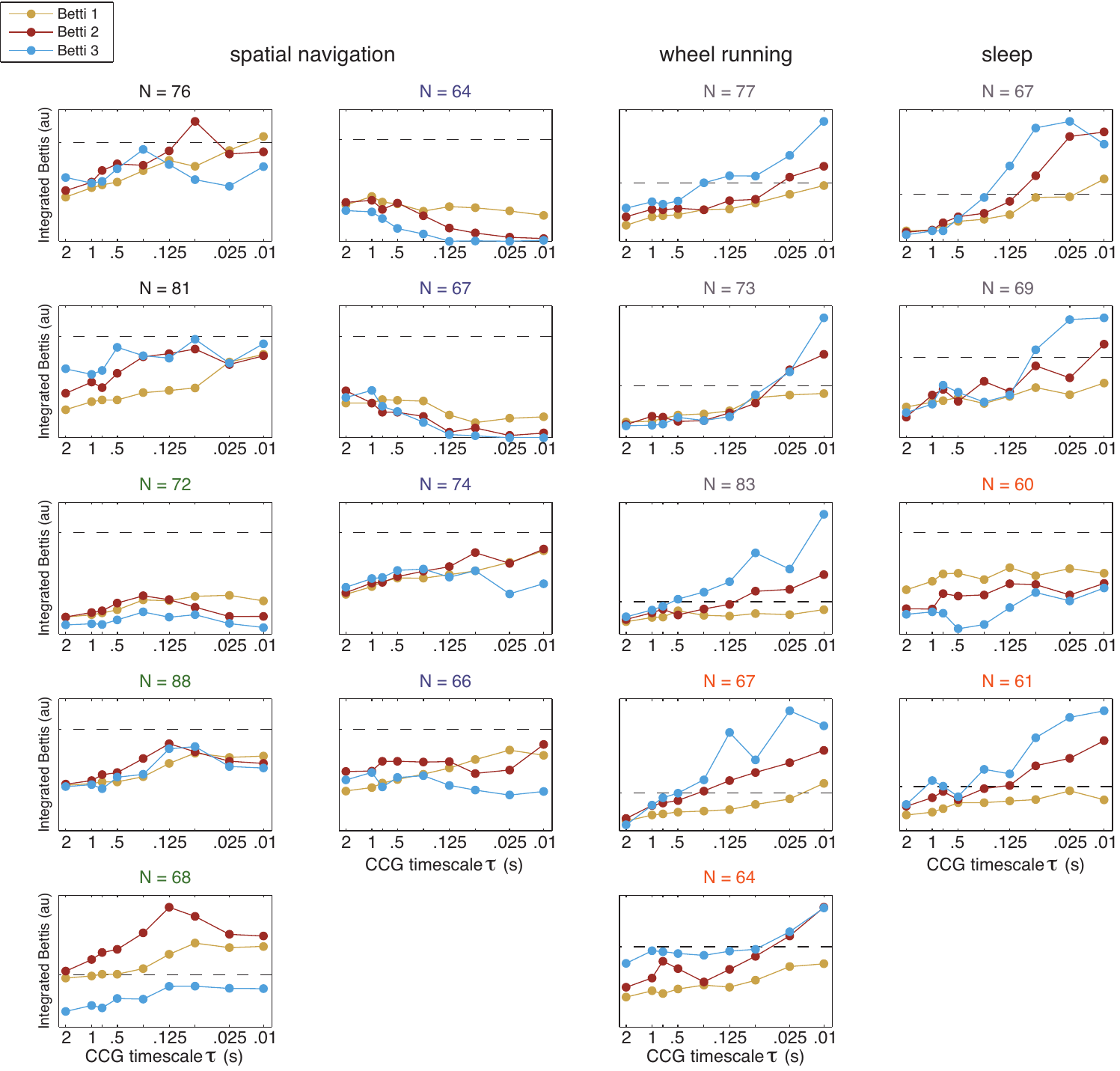}
 \end{center}

\vspace{-0.005in}

 {\bf Supplementary Figure 12:}  {\bf Integrated Betti values across a range of correlation timescales in spatial navigation, wheel running, and REM sleep data sets.}  
For each of the datasets, and each $m=1,2,3$, the integrated Betti number $\bar \beta_m =\int_0^1 \! \!\beta_m(\rho) d\rho$  was normalized  by its significance threshold   $b_m=Q_3+1.5\times(Q_3-Q_1)$ (see Supplementary Methods) that was obtained from the  distribution of the integrated geometric Betti curves $\bar \beta_m ^{\operatorname{geom}}=\int_0^1 \! \!\beta^{\operatorname{geom}}_m(\rho) d\rho$ in dimension $d=N$, where $N$ was the number of cells. 
 Each of the curves (yellow, red  and blue) correspond to the values of $\bar \beta_m / b_m $ in dimensions $m=1,2,3$ respectively. 
 The dashed line marks  the line  $\bar \beta_m   =b_m$;  the appropriate integrated Betti numbers were deemed consistent with geometric distribution if they lay below this line.  Note that the number $N$ of neurons is displayed in color (black, green, purple, gray and orange) to indicate recordings from the same animal; there were a total of 5 animals. 
 \end{figure}

 \newpage
 \begin{figure}[!h]\label{sfig13}
\begin{center}
\includegraphics[width=3.5 in]{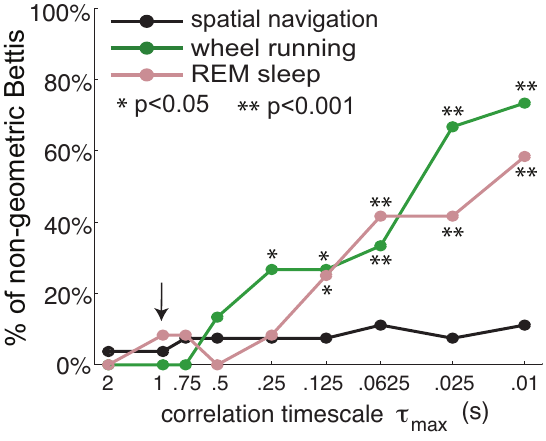}
 \end{center}

\vspace{-0.005in}

 {\bf Supplementary Figure 13:}    
 Percentage of  integrated Betti values that were {\it not} consistent with geometric controls (i.e. above the significance threshold) across all considered data sets as a function of the correlation timescale, $\tau_{\max}$, used to compute the pairwise correlation matrix (see Methods).  The arrow indicates the 1 s timescale used in the main figures.  All three behavioral conditions, spatial navigation (black), wheel running (green) and REM sleep (pink), are consistent with geometric structure at timescales ranging from .5 s to 2 s.  At finer timescales, however, the wheel running and REM sleep correlations are non-geometric, while the spatial navigation data remains consistent with geometric controls.

\bigskip 

For each timescale, we computed integrated Betti values $\bar{\beta}_1$, $\bar{\beta}_2$, and $\bar{\beta}_3$ for all data sets under three different conditions: 
(i) spatial navigation, (ii) wheel running, and (iii) REM sleep (see Supplementary Figure 12).  For each behavioral condition, we counted how many Betti values were above the significance threshold for rejecting the geometric hypothesis at each timescale.  Because our significance threshold rejects the geometric hypothesis at a rate of less than $5\%$, the $p$-value for a given condition and timescale satisfies
$$p < \sum_{\ell = k}^m {m \choose \ell} (0.05)^\ell (1-0.05)^{m-\ell},$$
where $k$ is the number of Betti values above the significance threshold, and $m$ is the total number of Betti values.
To obtain this upper bound on $p$-value, we used a binomial distribution with failure probability 0.05. Note that  this assumes Betti values are independent.  Although this is a reasonable assumption for Betti values from different data sets, Betti values from the same data set have statistical dependencies that are not well-understood.
 \end{figure}

 \end{document}